\documentclass{article}

\usepackage{arxiv}
\usepackage{url}
\usepackage{graphics}
\usepackage{epsfig}
\usepackage{color}
\usepackage{grffile}
\usepackage{subcaption}
\usepackage{hyperref}
\usepackage{multirow}
\usepackage{amsmath}
\usepackage{amsfonts}
\usepackage{array}
\usepackage{natbib}

\newcommand{\ignore}[1]{}

\newcolumntype{C}[1]{>{\centering\arraybackslash}p{#1}}

\newcommand{\typeset}[2]{#1^{\text{\tiny{#2}}}}
\newcommand{\ppair}[2]{\left<{#1},{#2}\right>}

\newcommand{\alignment}[3]{\left<\typeset{\mathcal{A}}{\text{\tiny #1}},\ppair{#2}{#3}\right>}

\newcommand{\alignmentsub}[4]{\left<\typeset{\mathcal{A}_{#1}}{\text{\tiny #2}},\ppair{#3_{#1}}{#4_{#1}}\right>}
\newcommand{\alignmentsubsst}[4]{\left<\typeset{\mathcal{A}_{#1}}{\text{\tiny #2}},\ppair{#3}{#4_{#1}}\right>}

\newcommand{\algn}[2]{\typeset{\mathcal{A}_{#1}}{#2}}

   \newcommand{\sr}[1]{\textcolor{red}{\textsc{SR:}#1}}

\title{The divergence time of protein structures modelled by Markov matrices and its relation to the divergence of sequences}

\author{
 $\text{Sandun Rajapaksa}^{1}$, $\text{Lloyd Allison}^{1}$, $\text{Peter J. Stuckey}^{1,2}$, $\text{Maria Garcia de la Banda}^{1}$, and $\text{Arun S. Konagurthu}^{1*}$ \\
  {1} Department of Data Science and Artificial Intelligence, Faculty of Information Technology\\
  Monash University, Clayton, VIC 3800, Australia \\
  {2} OPTIMA ARC Industrial Training and Transformation Centre \\
}

\begin{document}
\maketitle

\begin{abstract}
A time-parameterized statistical model quantifying the divergent evolution of protein structures in terms of the patterns of conservation of their secondary structures is inferred from a large collection of protein 3D structure alignments. This provides a better alternative to time-parameterized \emph{sequence}-based models of protein relatedness, that have clear limitations dealing with twilight and midnight zones of sequence relationships. Since protein structures are far more conserved due to the selection pressure directly placed on their function, divergence time estimates can be more accurate when inferred from structures.
We use the Bayesian and information-theoretic framework of Minimum Message Length to infer a time-parameterized stochastic matrix (accounting for perturbed structural states of related residues) and associated Dirichlet models (accounting for insertions and deletions during the evolution of protein domains). These are used in concert to estimate the Markov time of divergence of tertiary structures, a task previously only possible using proxies (like RMSD). By analyzing one million pairs of homologous structures, we yield a relationship between the Markov divergence time of structures and of sequences. Using these inferred models and the relationship between the divergence of sequences and structures, we demonstrate a competitive performance in secondary structure prediction against neural network architectures commonly employed for this task.
The source code and supplementary information are downloadable from \url{http://lcb.infotech.monash.edu.au/sstsum}
\end{abstract}

\section{Introduction} \label{sec:intro}

The evolutionary distance between two species is proportional to some (unknown) function of the time of divergence from their common ancestor. One way to estimate this time is by comparing the underlying macromolecular sequences that cascade the information of accumulated evolutionary changes across DNA$\rightarrow$RNA$\rightarrow$Proteins (sequence$\rightarrow$structure$\rightarrow$function). Since the introduction of the \emph{molecular evolutionary clock} by \citet{zuckerkandl1965evolutionary} to perform phylogenetic studies, several statistical models have been proposed to estimate the divergence of extant sequences from common ancestors, and to correlate the estimates of time from other sources of information (e.g., fossil records) when they exist~\citep{sarich1967immunological}. Such divergence time estimates require reliable statistical models of DNA/RNA/Proteins macromolecules~\citep{bromham2003modern}.

For protein amino acid sequences, several statistical models have been proposed to explain sequence variation as a function of time. The point accepted mutation (PAM) matrix of \citet{dayhoff1978} was the first successful model to explain the mutability of amino acid sequences. PAM is a stochastic (Markov) matrix defined in PAM (time) units where PAM-1 is a Markov matrix that embodies a 1\% expected change to the amino acids. Subsequent studies highlighted the importance of incorporating evolutionary time-dependent substitution and gap models as an elegant way to model the divergent relationships of proteins \citep{holmes1998studies,gonnet1992exhaustive}. The recent approach of \citet{sumanaweera2022bridging} derives a unified statistical model for quantifying the evolution of pairs of protein sequences using a time-parameterized Markov matrix (MMLSUM) for substitution events and associated time-parameterized Dirichlet model for insertion and deletion events.

Although deciphering sequence relationships and estimating their time of divergence is very useful, recent studies have shown clear limits of inference ~\citep{rajapaksa2022reliability}. When amino acid sequences diverge into the `twilight zone' of sequence relationships the quality of the relationships, and hence the estimates of time based on the amino acid models become unreliable.

Since protein structures are far more conserved than their sequences (due to the pressure of selection directly placed on protein function, derived from the structure) ~\citep{chothia1986relation,kinch2002evolution,pal2006integrated,konagurthu2006mustang,worth2009structural,soskine2010mutational,echave2016causes}, they yield better clues to infer reliable relationships and hence time estimates. Yet, in contrast to the statistically-rigorous models to compare amino acid sequences, there remains a dearth of reliable statistical models that can explain the observed divergence of  structures of homologous proteins and
correlate them with the corresponding models of amino acid evolution. In the absence of a statistical framework, these divergences are measured  by correlating the changes to the root-mean-square-deviation (RMSD) of main chain atoms (often only central $\alpha$-Carbon) compared to the proportion of amino acids that have change~\citep{chothia1986relation}.

The work presented here specifically aims to address the shortcoming of a statistical model to estimate the time of divergence between structures. Building on the Bayesian approach of Minimum Message Length inference~\citep{wallace2005statistical,allison2018coding}, we  infer and report a stochastic matrix and associated     Dirichlet models to estimate the Markov time of divergence of tertiary structures in terms of the patterns of conservation of their     resultant secondary structural states.  We refer to the resultant time-parameterized statistical models as SSTSUM.

Further, by analyzing a set of one million randomly chosen pairs of
protein domains related at the levels of `family' and `superfamily' derived from SCOPe~\citep{murzin1995scop}, we correlate the time of divergence of sequences with structures and establish a relationship.
Finally, as an application of the SSTSUM models and the relationship we could establish between the divergence of sequences and structures, we construct a classical statistical framework for the prediction of secondary structural states of amino acids. We then  compare this framework against three  neural network methods which demonstrate a competitive performance.

\section{Methods} \label{sec:methods}

\subsection{Minimum Message Length Framework} \label{sec:mmlframe}

The Minimum Message Length (MML) criterion provides an information-theoretic framework applicable to a large class of inference problems that involve hypothesis/model selection and parameter estimation. According to the  MML principle, the suitability of any hypothesis $H$ describing observed data $D$ can be quantified by measuring the Shannon information content $I(\cdot)$ as:
\begin{equation}\label{eq:mml}
I(H,D) = I(H) + I(D|H)
\end{equation}
The above formulation can be viewed as a lossless communication between an imaginary sender-receiver pair, where the sender losslessly encodes the hypothesis $H$ (taking $I(H) \text{ bits})$ followed by the data $D$ given the stated hypothesis $H$ (taking $I(D|H) \text{ bits})$.  This provides an objective trade-off between the complexity of the hypothesis, as captured by the first term $I(H)$, and its fidelity in explaining the observed data, as captured by the second term $I(H|D)$. Based on this principle, the best hypothesis ($H^*$) is the one that gives the shortest two-part message to communicate the observed data $D$. Further, implicit within the MML framework is a natural \emph{null} hypothesis test which encodes the observed data $D$ \emph{as is}. A  hypothesis $H$ is chosen only when it beats (i.e., is more succinct than) the message length of the \emph{null} model (i.e, $I(H, D) < I_{\text{NULL}}(D)$).

\subsection{Time-parameterized models for structural evolution} \label{sec:ssemodel}
Denote the primary, secondary, and tertiary information of any protein $S$  as $\{\typeset{S}{1D}, \typeset{S}{2D}, \typeset{S}{3D}\}$ respectively.  For a pair of proteins $\ppair{S}{T}$, let $\alignment{3D}{S}{T}$ represent an \emph{alignment} (residue-residue correspondence) derived by comparing their tertiary structures $\typeset{S}{3D}$ and $\typeset{T}{3D}$. We note that any alignment can be represented as a three-state string over the $\{\mathtt{match},
\mathtt{insert}, \mathtt{delete}\}$ states. Further, the residue-residue correspondence implicit in any alignment can be mapped across all levels (1D, 2D, 3D) of protein description. Thus, let $\alignment{3D$\mapsto$2D}{S}{T}$ imply the mapping of the residue-residue correspondences from a  structure alignment to their secondary structural states. Intuitively, this corresponds to the alignment of the secondary structural states of $\typeset{S}{2D}$ and $\typeset{T}{2D}$,
where the alignment relationship (of correspondences) is derived by aligning their tertiary structures.

Consider a large collection of structure alignments between homologous pairs of proteins denoted as:
$$
\typeset{\mathbf{D}}{3D}=\left\{\alignmentsub{1}{3D}{S}{T}, \ldots, \alignmentsub{|\mathbf{D}|}{3D}{S}{T}\right\}
$$
Implicit in this set of alignments is their mapping to the respective correspondences of secondary structural states, denoted as:
$$
\typeset{\mathbf{D}}{3D$\mapsto$2D}=\left\{\alignmentsub{1}{3D$\mapsto$2D}{S}{T}, \ldots, \alignmentsub{|\mathbf{D}|}{3D$\mapsto$2D}{S}{T}\right\}
$$

Given the dataset $\typeset{\mathbf{D}}{3D$\mapsto$2D}$ the main goal of this work is to infer the following two joint sets of statistical models:
\begin{enumerate}
    \item A time ($\typeset{t}{3D$\mapsto$2D}$) parameterized Markov matrix $\mathbf{M}$ modelling the divergence of correspondences defined by the alignments $\typeset{\mathbf{D}}{3D$\mapsto$2D}$.
    \item A corresponding time ($\typeset{t}{3D$\mapsto$2D}$) parameterized Dirichlet parameter $\pmb{\alpha}$ that works in concert with $\mathbf{M}$ to model the time-dependent 3-state machine parameters $\{\vec{\Theta}_i\}_{\forall 1\le i\le|\mathbf{D}|}$ (over  $\{\mathtt{match},\mathtt{insert}, \mathtt{delete}\}$ states) necessary to explain any alignment 3-state string $\algn{i}{3D$\mapsto$2D} \in \typeset{\mathbf{D}}{3D$\mapsto$2D}$.
\end{enumerate}
Note that the respective times of divergence $\{\typeset{t_1}{3D$\mapsto$2D},\ldots,\typeset{t_{|\mathbf{D}|}}{3D$\mapsto$2D}\}$ between pairs of proteins in dataset $\typeset{\mathbf{D}}{3D$\mapsto$2D}$ are \emph{unknown} and have to be estimated as a part of the inference.

To achieve our goal we use the MML principle, where it can be shown that the message length to  state the hypothesis $H=\{\mathbf{M},\pmb{\alpha},\{\typeset{t_i}{3D$\mapsto$2D},\Theta_i\}_{\forall 1\le i\le |\mathbf{D}|}\}$ is given by:
\begin{equation}\label{eq:firstpart}
    I(H) = I(\mathbf{M})+  I(\pmb{\alpha}) + \sum_{i=1}^{|\mathbf{D}|} (I(\typeset{t_i}{3D$\mapsto$2D}) + I(\vec{\Theta}_i|\pmb{\alpha}(\typeset{t_i}{3D$\mapsto$2D}))) \qquad \text{bits}
\end{equation}
and the message length to state all the alignment information in the dataset $\typeset{\mathbf{D}}{3D$\mapsto$2D}$ using the above hypothesis is given by:
\begin{equation}
\begin{split} \label{eq:secondpart}
    I(\typeset{\mathbf{D}}{3D$\mapsto$2D}|H) =  \sum_{i=1}^{|\mathbf{D}|} & (I(\algn{i}{3D$\mapsto$2D}| \vec{\Theta}_i) \\
& + I(\ppair{\typeset{S_i}{2D}}{\typeset{T_i}{2D}} | \algn{i}{3D$\mapsto$2D},\mathbf{M}(\typeset{t_i}{3D$\mapsto$2D}))) \qquad \text{bits} \end{split}
\end{equation}

Combining Equations \ref{eq:firstpart} and \ref{eq:secondpart} gives the total message length objective that we want to minimize:
\begin{equation}\label{eq:objfunc}
H^{*}= \arg\min_{\forall H} I(H) + I(\typeset{\mathbf{D}}{3D$\mapsto$2D}|H) \quad\text{bits}
\end{equation}

Each term on the right-hand sides of  Equations \ref{eq:firstpart} and \ref{eq:secondpart} is explained in detail below.

\paragraph{\textbf{$I(\mathbf{M})$ term: }}
$\mathbf{M}$ is a stochastic Markov matrix in its base-matrix form over $K$ secondary structural states.  Here, $K=3$ which  accounts for the secondary structural categories of $\{ \text{Helix}~\mathtt{H}, \text{Strand}~\mathtt{E}, \text{Coil}~\mathtt{C}\}$. Each column vector in this matrix is $\mathbb{L}_1$-normalized ($\sum_{j=1}^K \mathbf{M}_{i,j} = 1, \forall 1\le i\le K$) and represents a set of points in a $(K-1)$-simplex. $\mathbf{M}(1)$ or $\mathbf{M}$ refers to the \emph{base matrix}
which corresponds to the transition probabilities between any two states at the Markov divergence time $\typeset{t}{3D$\mapsto$2D} = 1$. This is calibrated such that there is a 1\% expected change of those states (and this is merely a convention to define a unit of Markov time under this model). It is the property of a Markov matrix that any $\mathbf{M}(\typeset{t}{3D$\mapsto$2D})$ can be derived by exponentiating $\mathbf{M}$ by $\typeset{t}{3D$\mapsto$2D}$: $\mathbf{M}(\typeset{t}{3D$\mapsto$2D}) = \mathbf{M}^{\typeset{t}{3D$\mapsto$2D}}$. We note that the expected change of a
stochastic matrix $\mathbf{M}(\typeset{t}{3D$\mapsto$2D})$ gives the expected proportion of substitutions to be observed if data (alignments) followed that model. Since any diagonal cell of  the matrix represents the probability of observing the conservation of the secondary structural state (i.e., no change to the state), the weighted average (using the background probability of each state, derived from the stationary distribution of $\mathbf{M}$) gives the expected probability of secondary
structure states being conserved under the model. Hence, the expected change implicit in the state of the Markov matrix $\mathbf{M}$ at any time $\typeset{t}{3D$\mapsto$2D}$ is calculated as:
\begin{equation}\label{eq:expchange}
    \text{Expected change of } \mathbf{M}(\typeset{t}{3D$\mapsto$2D}) = 1.0 - \sum_{i=1}^{K} \pi_i \mathbf{M}^{\typeset{t}{3D$\mapsto$2D}}_{i,i}
\end{equation}
where $\pi_i$ is the stationary probability of each secondary structure state and $\mathbf{M}^{\typeset{t}{3D$\mapsto$2D}}_{i,i}$ is the probability on each diagonal cell of the stochastic Markov matrix $\mathbf{M}$ at time ${\typeset{t}{3D$\mapsto$2D}}$. The eigen decomposition of the stochastic matrix $\mathbf{M}$ enables an efficient computation of $\mathbf{M}(\typeset{t}{3D$\mapsto$2D})$ where $\mathbf{M}^{\typeset{t}{3D$\mapsto$2D}} = \mathbf{S}\Lambda^{\typeset{t}{3D$\mapsto$2D}}\mathbf{S}^{-1}$, $\mathbf{S}$ is
the eigenvector and $\Lambda$ is the diagonal eigenvalue matrix of $\mathbf{M}$. Therefore, when the base matrix is inferred once, the  matrix at time ${\typeset{t}{3D$\mapsto$2D}}$  can be computed by three $K\times K$ matrix multiplications, after raising each of the $K$ eigenvalues to the power ${\typeset{t}{3D$\mapsto$2D}}$. Hence, we only need to state the base matrix $\mathbf{M}$ under the Shannon information term $I(\mathbf{M})$ in Equation \ref{eq:firstpart}. The message length formulation of this term is given in supplementary section S2.

\paragraph{$I(\pmb{\alpha}), I(\typeset{t_i}{3D$\mapsto$2D}), I(\vec{\Theta}_i|\pmb{\alpha}(\typeset{t_i}{3D$\mapsto$2D})) $ \textbf{terms: }}
Any alignment relationship  $\algn{i}{3D$\mapsto$2D}$ of any protein pair $\ppair{S_i}{T_i}$ that have diverged over time $\typeset{t_i}{3D$\mapsto$2D}$ can be represented as a 3-state string over $\{\mathtt{match},\mathtt{insert},\mathtt{delete}\}$ states. Hence, the encoding of  $\algn{i}{3D$\mapsto$2D}$ is carried out using a multi-state distribution estimating the 3-state machine parameters $\Theta_i$ using time-parameterized Dirichlet priors $\mathtt{Dir}(\vec{\alpha})$ with a parameter vector $\vec{\alpha} = [\alpha_1,\alpha_2, \ldots,\alpha_p]$.  Each $\vec{\alpha}$ parameter vector  models multi-state probability vector $\vec{\Theta}_i$  in a ($p-1$)-simplex space.  The optimal estimates of time-parameterized $\pmb{\alpha}$ are searched together with the Markov matrix $\mathbf{M}$, by minimizing the two-part message length, as per the objective in Equation \ref{eq:objfunc}. (See supplementary section S2.)

In Equation \ref{eq:firstpart} the term $I(\pmb{\alpha})$ is the statement length of the set of time-dependent Dirichlet parameters $\pmb{\alpha}$; The term $I(\typeset{t_i}{3D$\mapsto$2D})$ is the encoding length of the inferred optimal time of divergence $\typeset{t_i}{3D$\mapsto$2D}$ using $\left<\typeset{S_i}{2D},\typeset{T_i}{2D}\right>$ given its alignment $\algn{i}{3D$\mapsto$2D}$  (see Section \ref{sec:opttime}). The term $I(\vec{\Theta}_i|\pmb{\alpha}(\typeset{t_i}{3D$\mapsto$2D}))$ is the encoding length of the free parameter vector $\vec{\Theta}_i$ of the 3-state alignment machine parameters inferred for each alignment $\algn{i}{3D$\mapsto$2D}\in \typeset{\mathbf{D}}{3D$\mapsto$ 2D}$. See supplementary section S2 for more details.

\paragraph{\textbf{Terms in Equation \ref{eq:secondpart}:}} $ I(\algn{i}{3D$\mapsto$2D}| \vec{\Theta}_i)$ is the encoding length of each three-state alignment string $\algn{i}{3D$\mapsto$2D}$; and  $I(\ppair{\typeset{S_i}{2D}}{\typeset{T_i}{2D}} | \algn{i}{3D$\mapsto$2D},\mathbf{M}(\typeset{t_i}{3D$\mapsto$2D}))$ is the encoding length of the message needed to explain all secondary structural states in each pair $\ppair{\typeset{S_i}{2D}}{\typeset{T_i}{2D}}$. For each alignment, $\algn{i}{3D$\mapsto$2D}$, $\mathbf{M}(\typeset{t_i}{3D$\mapsto$2D})$ is used to encode all $\texttt{matched}$ secondary structural states defined by that alignment, and the stationary distribution of $\mathbf{M}$ (which by definition is invariant with time) is used to encode \texttt{inserted} and \texttt{deleted} regions defined by that alignment. The computation of each term of this objective function is described in detail in the supplementary materials S2.

\subsection{Search for the best Markov matrix $\mathbf{M}^*$ and associated time-dependent Dirichlet parameters $\pmb{\alpha}$} \label{sec:searchmalpha}

Given the objective function in Equation \ref{eq:objfunc} and a dataset $\typeset{\mathbf{D}}{3D$\mapsto$2D}$, we can search for the best matrix $\mathbf{M}^*$ and associated Dirichlet parameters $\pmb{\alpha}$ (those that minimize the objective function) by employing an Expectation-Maximization (EM) like strategy over a simulated annealing search method. In general, EM-like techniques are often used to optimize the parameters of a statistical model with unknown  variables \citep{dempster1977maximum}.  In this work, the EM-like search involves holding $\pmb{\alpha}$ (and consequently $\vec{\Theta}$) fixed while performing a simulated annealing approach to search for the best matrix $\mathbf{M}$ and associated alignment time parameters. Then, the matrix $\mathbf{M}$ and time parameters  $\{\typeset{t_1}{3D$\mapsto$2D},\ldots,\typeset{t_{|\mathbf{D}|}}{3D$\mapsto$2D}\}$ are held fixed to estimate the best parameters $\pmb{\alpha}(\typeset{t_i}{3D$\mapsto$2D})$ and $\vec{\Theta}_i$. This process is repeated until the objective function converges. See supplementary materials S3 for the methodological details and initial parameters used in the inference process.

\subsection{Estimating the optimal time for a given structure alignment} \label{sec:opttime}

Given an alignment  $\{\alignmentsub{i}{3D$\mapsto$2D}{S}{T}\}\in\typeset{\mathbf{D}}{3D$\mapsto$2D}$,  $\{\typeset{t_i}{3D$\mapsto$2D}\}$ is inferred using the MML framework, by minimizing the two-part  encoding length of $I(\alignmentsub{i}{3D$\mapsto$2D}{S}{T})$, which can be estimated by:
\begin{small}
\begin{equation}\label{eq:objfuncalign}
\begin{split}
    I\left(\alignmentsub{i}{3D$\mapsto$2D}{S}{T}\right) &= I(\typeset{t_i}{3D$\mapsto$2D}) +  I(\vec{\Theta}_i|\pmb{\alpha}(\typeset{t_i}{3D$\mapsto$2D})) + I(\algn{i}{3D$\mapsto$2D} | \vec{\Theta}_i) \\&+  I(\ppair{\typeset{S_i}{2D}}{\typeset{T_i}{2D}} | \algn{i}{3D$\mapsto$2D}, \mathbf{M}(\typeset{t_i}{3D$\mapsto$2D})) \qquad \text{bits}
\end{split}
\end{equation}
\end{small}

\subsection{Function  model for secondary structure prediction}\label{sect:predframework}

This subsection is related not to inference but to one of the applications of the inferred Markov models that we demonstrate here: its use in secondary structure prediction.

Protein secondary structure prediction can be formulated as a function model where the output attributes depend on the input attributes. In general, the main aim of a fully parameterized function model is to provide the conditional probability of the output given the input, i.e $\Pr(H|D)$. Since the input data is common knowledge, a sender need not encode them in any message (i.e $\Pr(D) = 1$) \citep{allison2018coding}. Applying this insight to the Bayes theorem, we get $\Pr(H,D) = \Pr(H|D)$
(i.e, $I(H,D) = I(H|D)$). This gives rise to a statistically robust objective function to optimize which is to find the best hypothesis on the data that maximizes the probability of $H$ and $D$ or equivalently that minimizes the total message length $I(H, D)$. Our approach to structure prediction is in this class of statistical models. Here, one aims to make  a prediction of secondary structural states of protein from its amino acid sequence information. In other words, the prediction
is the hypothesis. The data/evidence it uses is a set of (non-redundant) proteins with known structures. Further, there is no `leakage'\citep{gibney2022ai} of information about the protein whose secondary structure is being predicted within this structural  data.

\begin{figure}[b]
        \centering
         \begin{subfigure}[b]{0.49\textwidth}
             \centering
             \includegraphics[width=\textwidth]{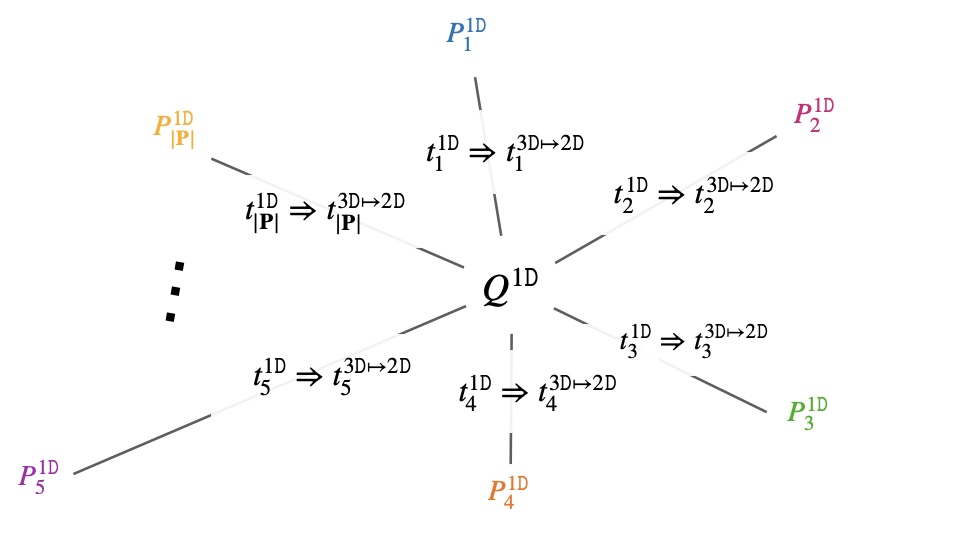}
             \caption{A star model of sequence relationship of $\typeset{Q}{1D}$ (center of star) with hits in $\mathbf{P}$ as its points.}
             \label{fig:starmodela}
         \end{subfigure}
         \begin{subfigure}[b]{0.49\textwidth}
             \centering
             \includegraphics[width=\textwidth]{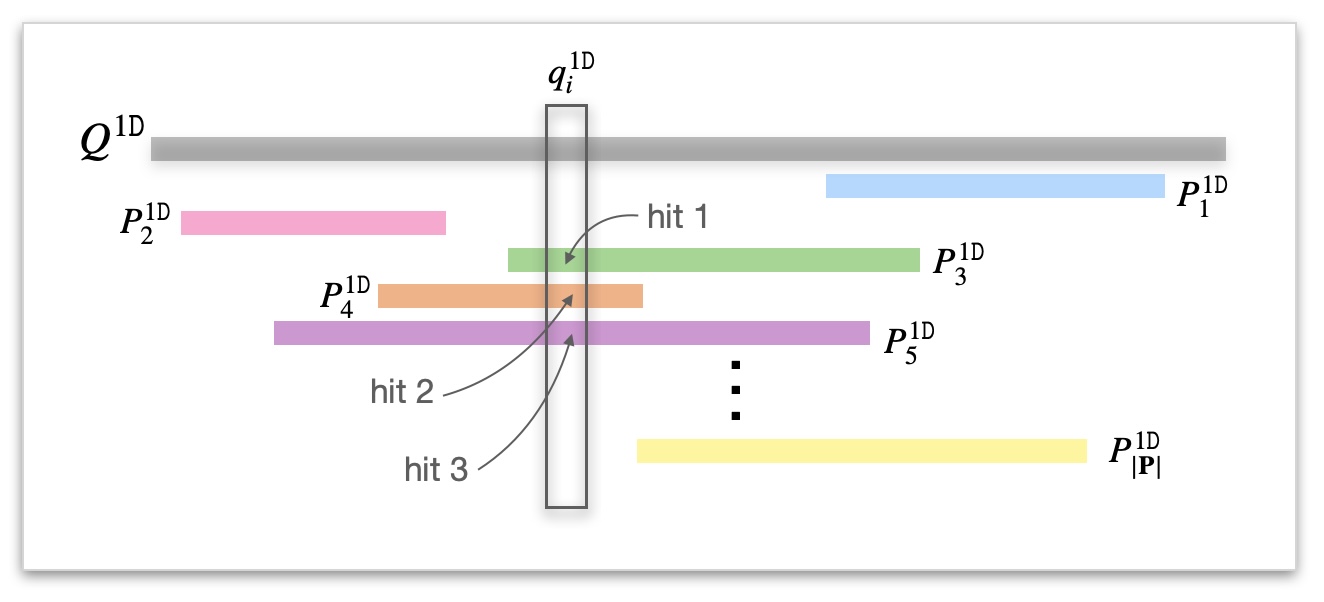}
             \caption{Visualization of a set of hits in $\mathbf{P}$ covering the $\typeset{q_i}{1D}$th position of $\typeset{Q}{1D}$.}
             \label{fig:starmodelb}
         \end{subfigure}
            \caption{}
	\label{fig:starmodel}
\end{figure}

\subsubsection{Combining multiple sources of evidence/hits}
For a given query amino acid sequence  $\typeset{Q}{1D} = \{\typeset{q_1}{1D}, \typeset{q_2}{1D}, \ldots \typeset{q_{|Q|}}{1D}\}$, we infer a set of \emph{hits} (local sequence alignments) with regions of proteins in the protein data bank: $\mathbf{P} = \{ \alignmentsubsst{1}{1D}{\typeset{Q}{1D}}{P}, \ldots, \alignmentsubsst{|\mathbf{P|}}{1D}{\typeset{Q}{1D}}{P}\}$. Associated with each hit is an optimal estimate of time $t_i^{\text{\tiny 1D}}$ obtained using any amino acid stochastic matrix (e.g.
MMLSUM~\citep{sumanaweera2019statistical,sumanaweera2022bridging}).

The set $\mathbf{P}$ along with the corresponding time $\{t_i^{\text{\tiny 1D}}\}$ parameters can be modeled over a star model of sequence relationships, with the query   $\typeset{Q}{1D}$ as the center (see Fig. \ref{fig:starmodela}).
Using this star model of relationship, we aim to estimate  the conditional probabilities for each amino acid  $\typeset{q_i}{1D}\in \typeset{Q}{1D}$ to be in any secondary structural state, by linearly combining their sources of evidence/hits in $\mathbf{P}$ (discussed below).

The innovation needed here is to correlate any estimated amino acid time of divergence $\typeset{t_i}{1D}$ with the time of divergence of structures $\typeset{t_i}{3D$\mapsto$2D}$. One of the contributions of this work is the inference of relationship $\typeset{t_i}{1D}\implies\typeset{t_i}{3D$\mapsto$2D}$, described in Results \ref{sec:seqstrtime}.

Let ~$\aleph^{\text{\tiny 2D}} = \{a_1,a_2,\ldots a_K\}$ define the alphabet of secondary structure states and $\typeset{\tilde{Q}}{2D} = (\typeset{\tilde{q}_1}{2D},\typeset{\tilde{q}_2}{2D},\ldots, \typeset{\tilde{q}_{|Q|}}{2D})$ denote the predicted secondary structural states of the amino acid sequence $\typeset{Q}{1D}$ that we aim to compute.  We combine the information of each hit in $\mathbf{P}$ to estimate the conditional probabilities of each $\typeset{q_i}{1D}$ being involved in any of the secondary structural states as follows. Refer to Fig.  \ref{fig:starmodelb}. For any specific amino acid  $\typeset{q_i}{1D}\in \typeset{Q}{1D}$, consider the subset $\mathbf{H}$ of local alignments in $\mathbf{P}$ (whose structures are known) that correspond to a \texttt{match} state with $\typeset{q_i}{1D}$ in their respective alignments denoted as $\mathbf{H}=\{\typeset{h_1}{1D},\typeset{h_2}{1D},\ldots \typeset{h_{|\mathbf{H}|}}{1D}\}$. Further, let the amount of compression gained by the subset of local alignments covering the $\typeset{q_i}{1D}$ amino acid be defined by a corresponding set of bit lengths $\{c_1,c_2,\ldots c_{|\mathbf{H}|}\}$.

Since the \emph{true} secondary structure state $\typeset{q_i}{2D}$ is unknown (and something we want to predict), the above star model allows us to  combine the evidence of all hits in $\mathbf{P}$ to derive the conditional probability estimates for each $\typeset{q_i}{1D}$ being in any of the three possible secondary structural states. Importantly,  the Markov matrix $\mathbf{M}$ and the relationship between $\typeset{t_i}{1D}\implies\typeset{t_i}{3D$\mapsto$2D}$ that we derive in this work
is used to compute these probabilities, as follows:

\begin{small}
\begin{equation}
\begin{split} \label{eq:predfunction}
    \Pr(\typeset{q_i}{2D} = a_j\in\typeset{\aleph}{2D} | \mathbf{P}) = & \frac{\sum_{k = 1}^{|\mathbf{H}|} \left(2^{c_k} \times  \Pr(a_j |\typeset{h_k}{2D},\mathbf{M},\typeset{t_i}{3D$\mapsto$2D})\right)}{\sum_{l=1}^{|\mathbf{H}|} 2^{c_l}}  , \\
& \forall 1\le i\le |Q|, 1\le j\le |\aleph^{\text{\tiny 2D}}| \end{split}
\end{equation}
\end{small}
The term $\frac{2^{c_k}}{\sum_{l=1}^{|\mathbf{H}|} 2^{c_l}}$ defines the weight/probability derived from the extent of compression (measured in bits) given by local alignments covering $\typeset{q_i}{1D}$, where $2^{c_k}$ gives the \emph{odds} of the sequence relationship before normalization. Most importantly, the estimate of
$ \Pr(a_j |\typeset{h_k}{2D},\mathbf{M},\typeset{t_i}{3D$\mapsto$2D})$ comes from the inferred stochastic matrix $\mathbf{M}$  at time  $\typeset{t_i}{3D$\mapsto$2D}$.

\subsubsection{The best secondary structure prediction}

After computing the conditional  probability of assignment of a secondary structure state at any position in the query amino acid sequence $\typeset{Q}{1D}$ over all possible secondary structure states, the best prediction of $\typeset{\tilde{q}_i}{2D}\in\typeset{Q}{2D}$ is chosen as:
$$
\typeset{\tilde{q}_i}{2D} = \arg\max_{j=1}^{|\aleph^{\text{\tiny 2D}}|}
\Pr(\typeset{q_i}{2D}=a_j\in\typeset{\aleph}{2D} | \mathbf{P})
$$

\ignore{ Here the second and third terms can be computed using the null probabilities in $\mathbf{N}$. Since $\mathcal{A}$ is a sequence alignment and the secondary structure assignment of $S$ is the prediction and hence unknown, we estimate the optimal secondary structure time $t$ by first computing the optimal amino acid time using SeqMMLigner  \citep{sumanaweera2019statistical} and then mapping it to the secondary structure time. \sr{see section}. Then we compute the first term in Equation \ref{eqn:alignprob} using the corresponding $\mathbf{M}(t)$, considering all possibilities for $\mathtt{SSE}(s_i)$ where  $\mathtt{SSE}(s_i)\in \mathbf{Q}$. }

\section{Results and Discussion} \label{sec:results}

Using the MML inference described above, we inferred a time-parameterized stochastic Markov matrix $\mathbf{M}$ and associated Dirichlet parameters $\pmb{\alpha}$ using the secondary structure alphabet $\aleph^{\text{\tiny 2D}}  = \{\mathtt{H}~\text{(Helix)}, \mathtt{E}~\text{(Extended strand of sheet)}, \mathtt{C}~\text{(Coil)}\}$. We term this time-parameterized model inferred in this work, SSTSUM. The inference was based on the
same benchmark dataset $\mathtt{SCOP2}$ \citep{sumanaweera2022bridging} which was used to infer MMLSUM (a time-parameterized model for amino acid substitutions). This dataset contains 59,092 unique domain pairs sampled from family and superfamily levels of SCOP (v2.07) \citep{murzin1995scop}.

\begin{figure}
    \centering
    \includegraphics[width=0.45\textwidth]{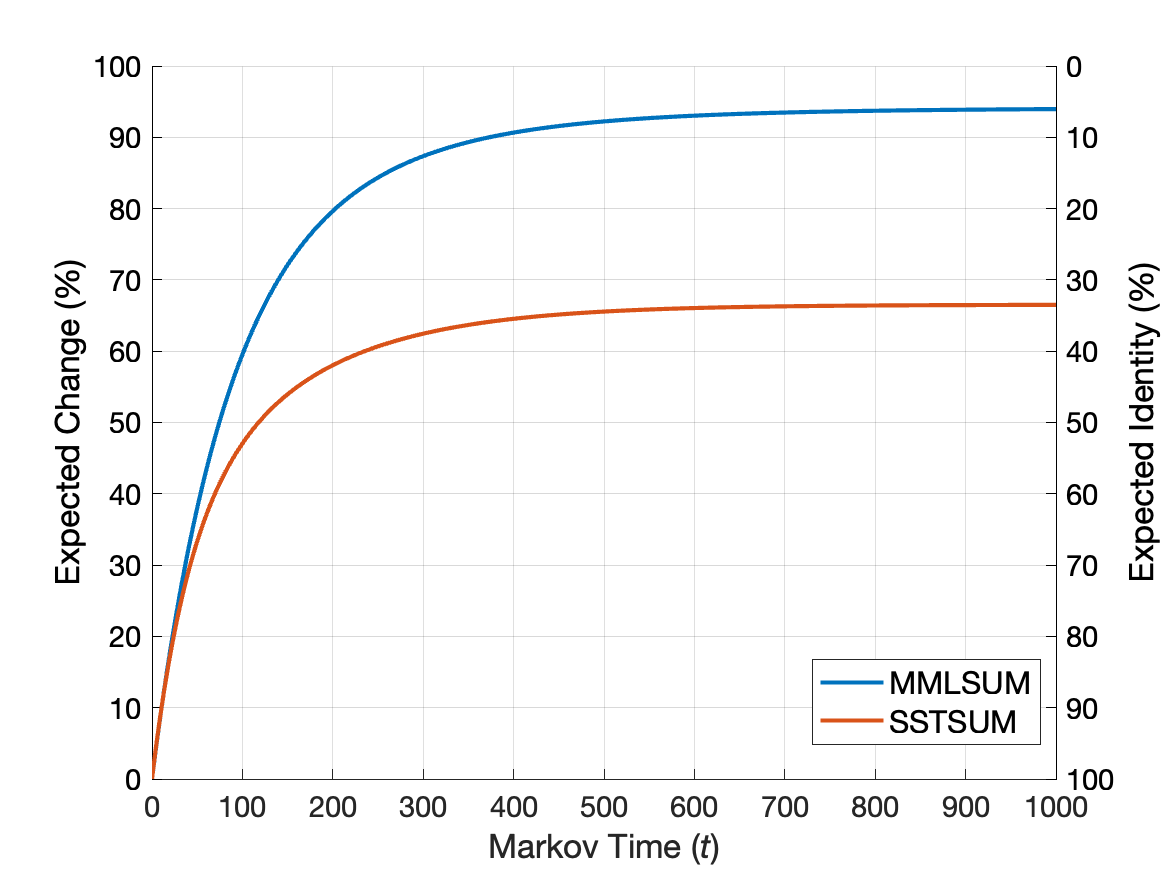}
    \caption{Mathematical expectation of the extent of change implied by the SSTSUM model (over secondary structure states) compared to MMLSUM (over amino acids), as a function of time.}
    \label{fig:expchange}
\end{figure}

SSTSUM is represented canonically in its base matrix form $\mathbf{M}$ at time $t=1$, modeling 1\% expected changes at the level of secondary structures. To calculate the state of the stochastic matrix at time $t$ requires an exponentiation of the base matrix by $t$,  $\mathbf{M}^t$. We emphasize that the unit of time is internal to any stochastic matrix. For example, the unit of time for MMLSUM stochastic matrix (over 20 amino acid states) yielding 1\% expected change is not the same
as the unit of time for SSTSUM (over 3 secondary structure states) as mentioned before. At the outset, the  relationship between the two notions of time is unknown and yet to be established. This is explored in the following analysis.

First, we start by plotting the expected change of secondary structural states and amino acid states on the same Markov time axis (see Fig. \ref{fig:expchange}).
We note that the expected change of SSTSUM at any time $t$ ($\mathbf{M}^t$) is computed using Equation \ref{eq:expchange} (see Section \ref{sec:ssemodel}). A similar computation is involved in calculating the expected change for MMLSUM over 20 amino acid states.

From this initial figure, we can gather that  an expected change of 60\% (40\% identity) is reached at Markov time $t=236$ for SSTSUM but at  $t=102$ for MMLSUM. Similarly, at $t=400$, the expected change is about 65\%  for SSTSUM and about 90\% for MMLSUM. In the limit, both stochastic matrices converge to their respective stationary distributions, with the expected change of 66.5\% (for SSTSUM) and 94.1\% (for MMLSUM).  To properly understand
the relationship between these two notions of time, we conduct the following comprehensive comparison between the time of divergence of sequences and  of structures.

\subsection{Divergence time of structures ($\typeset{t}{3D$\mapsto$2D}$) and its relation to sequences ($\typeset{t}{1D}$)} \label{sec:seqstrtime}
\begin{figure*}[htb] \begin{tabular}{@{}c@{}c@{}c@{}} \includegraphics[width=0.33\textwidth]{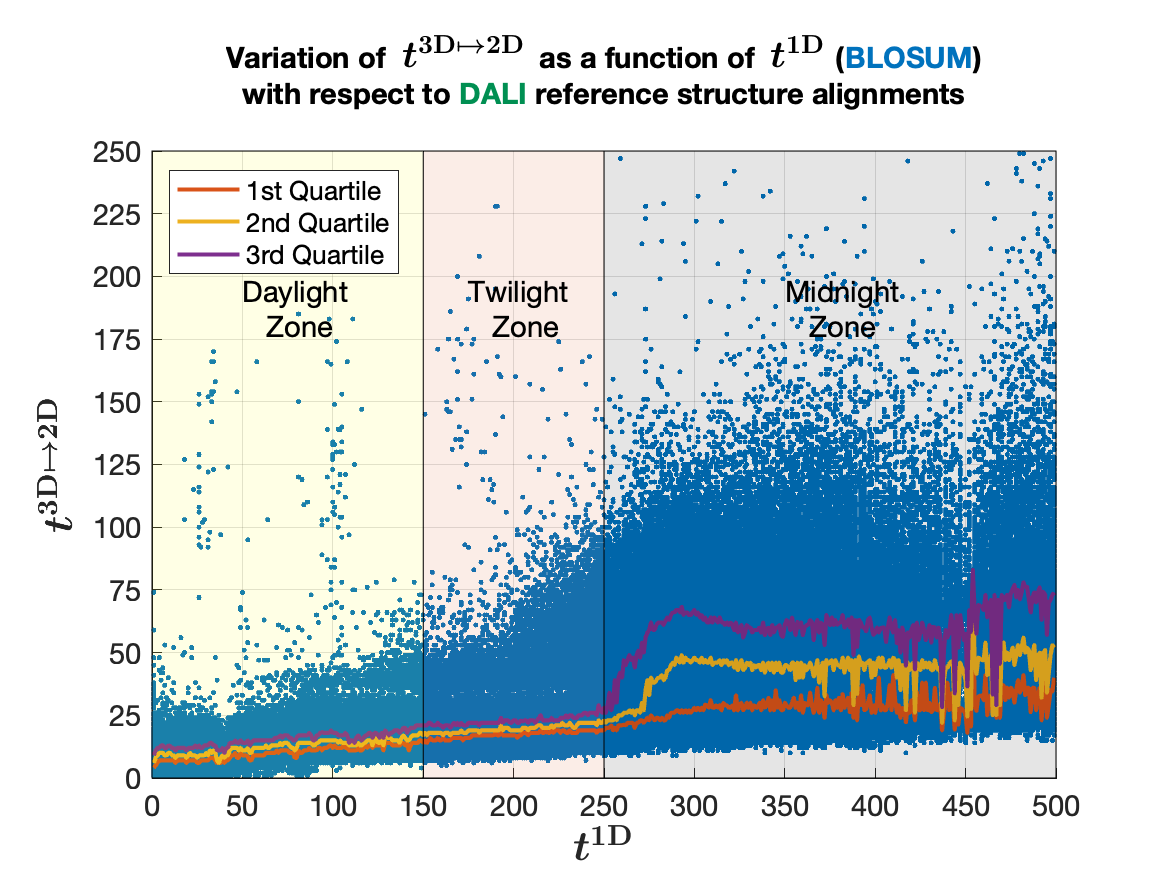} & \includegraphics[width=0.33\textwidth]{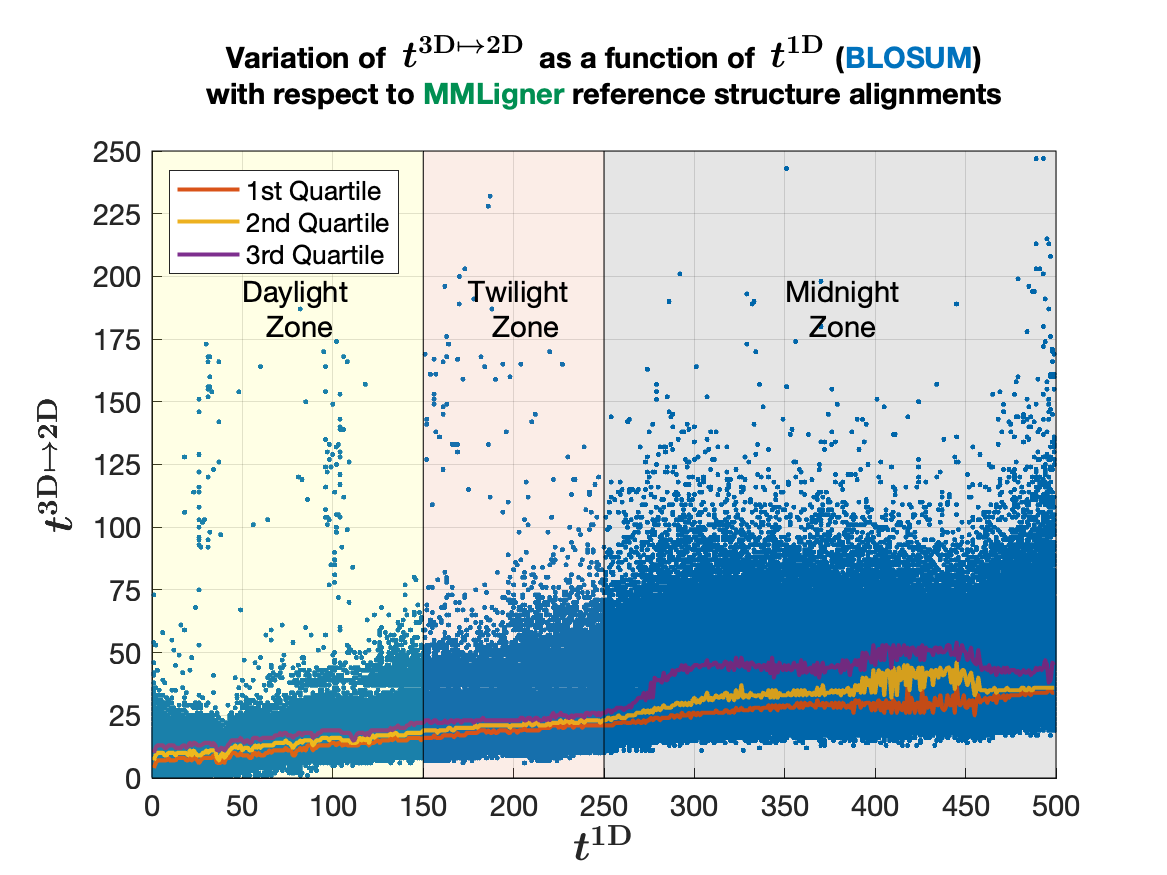} & \includegraphics[width=0.33\textwidth]{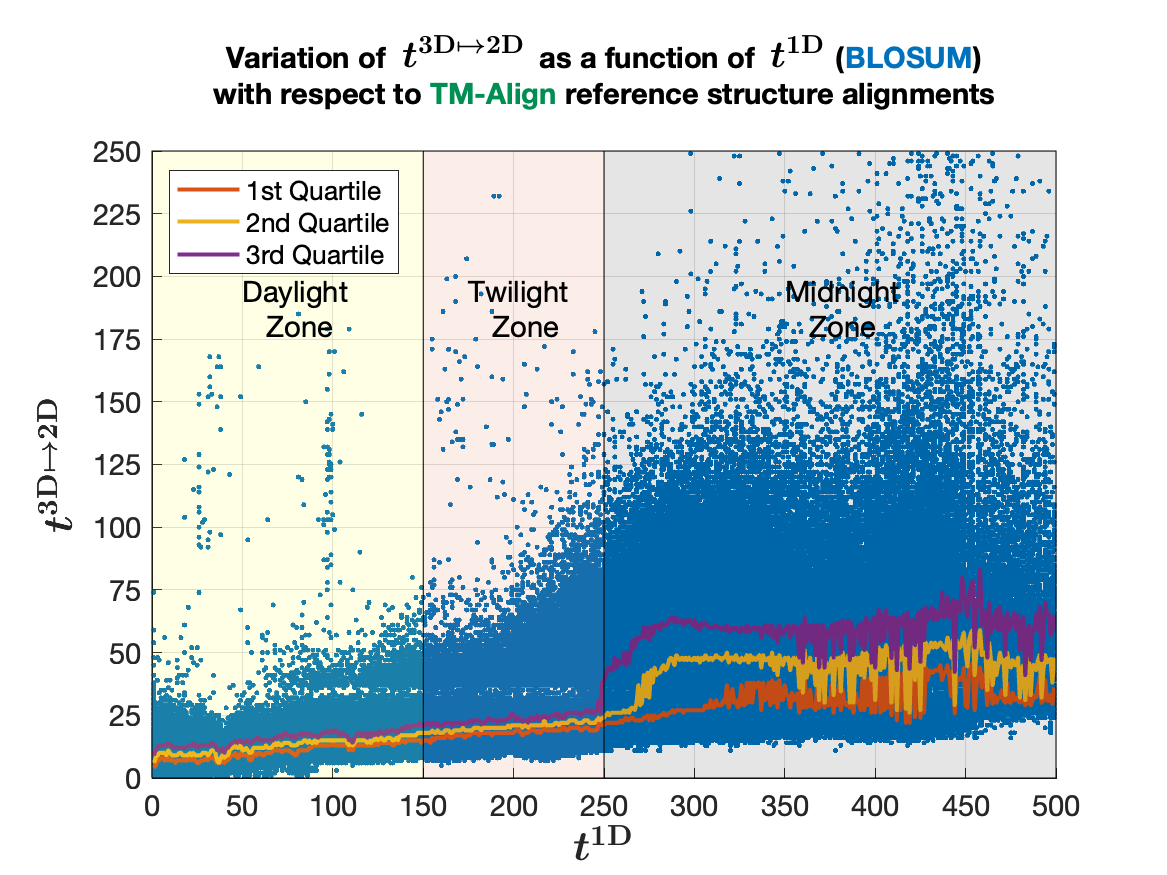} \\ \includegraphics[width=0.33\textwidth]{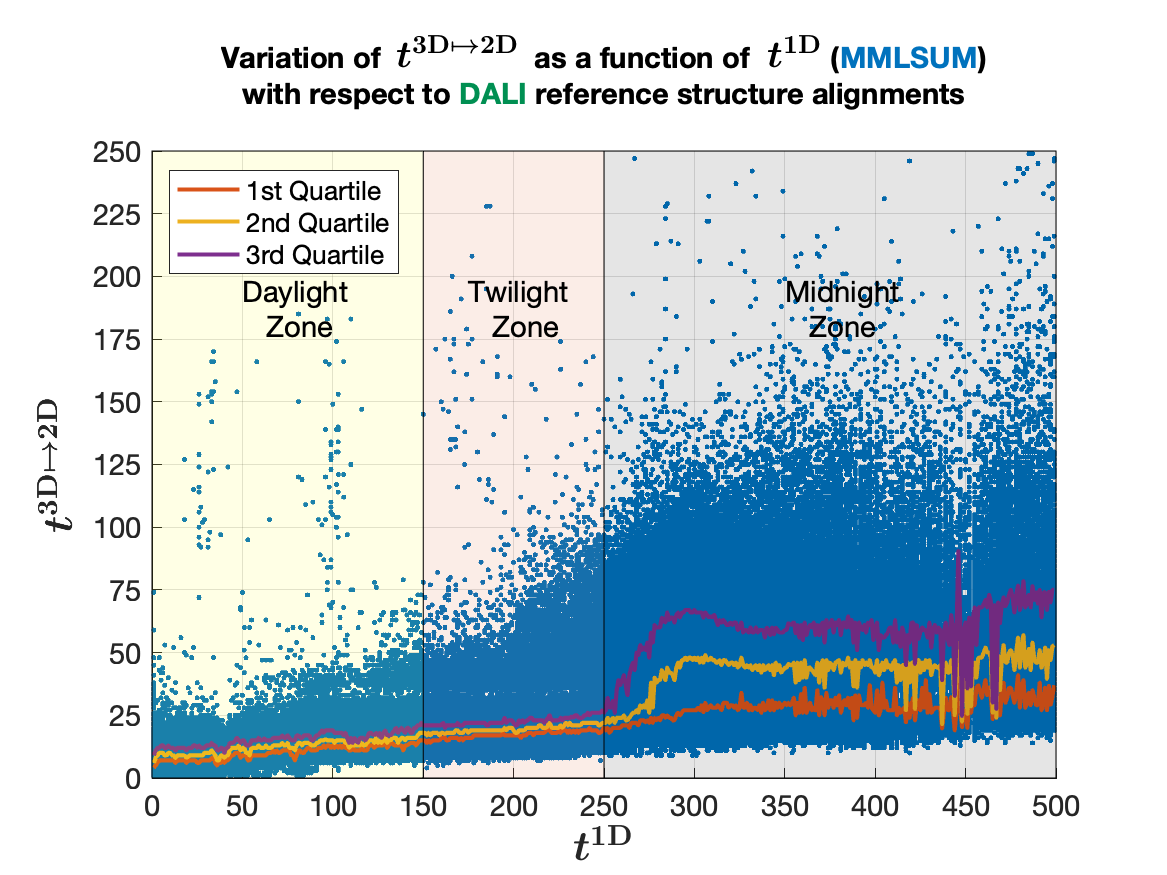} &
\includegraphics[width=0.33\textwidth]{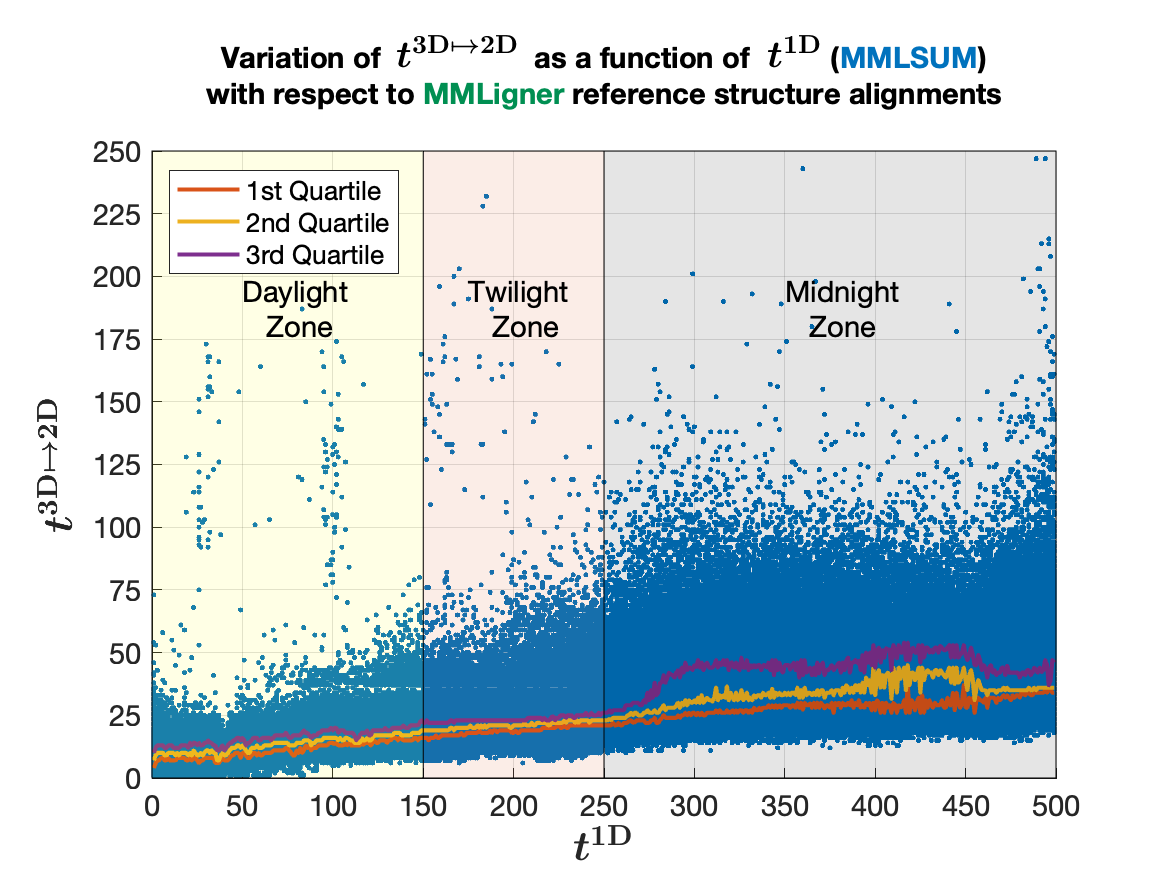} & \includegraphics[width=0.33\textwidth]{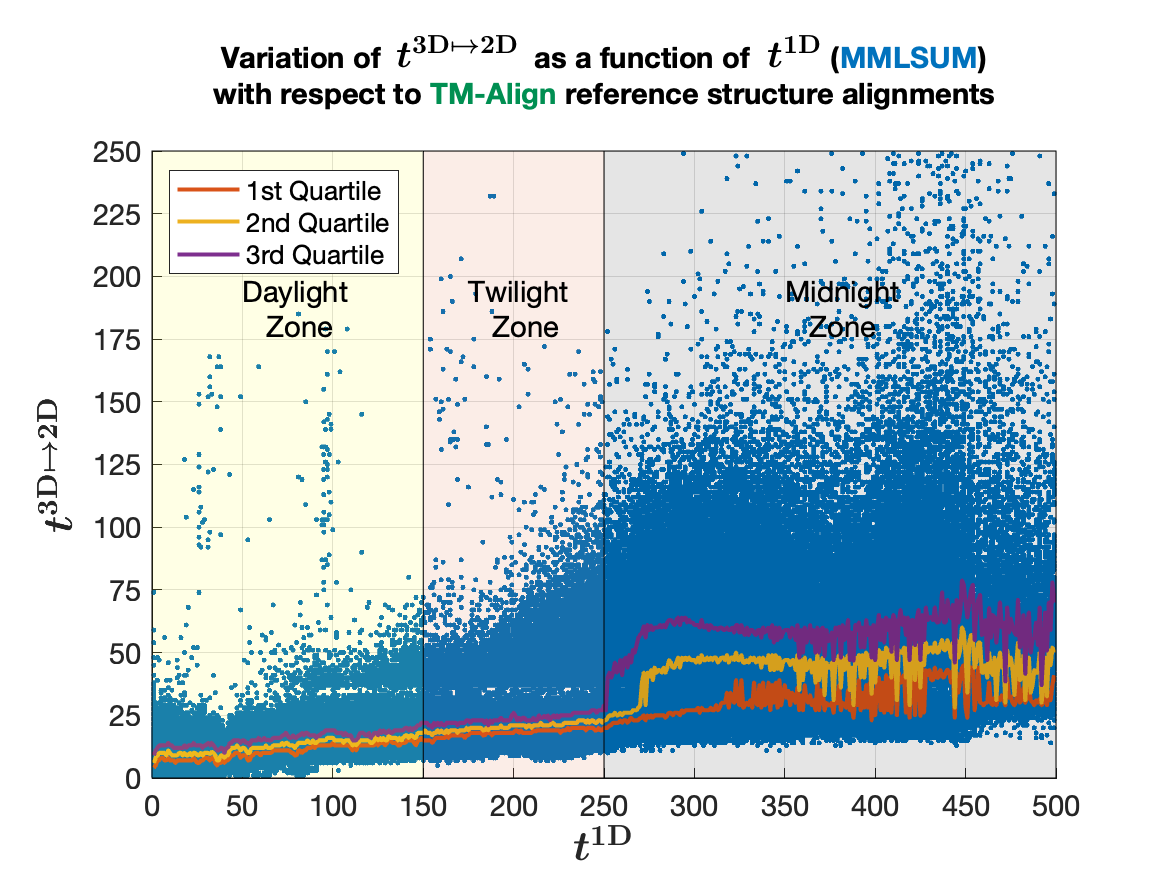} \\ \includegraphics[width=0.33\textwidth]{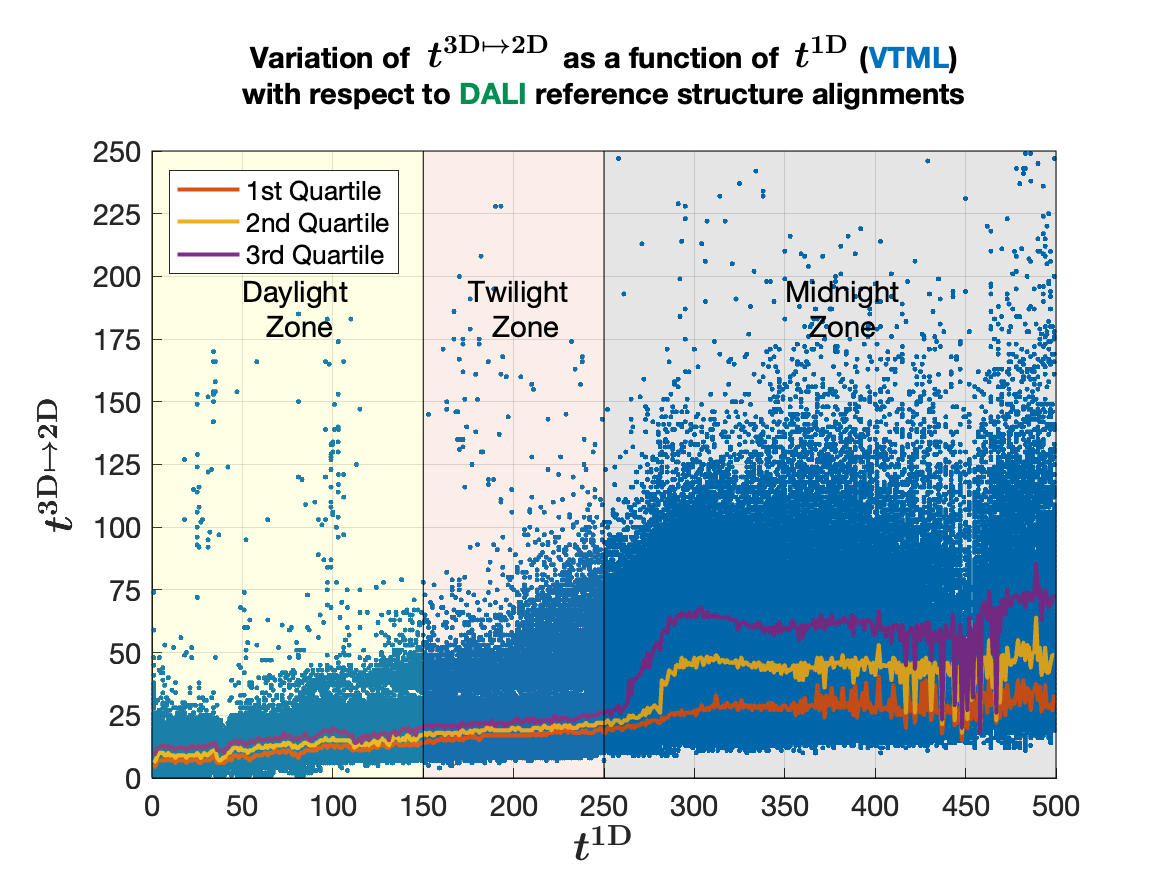} & \includegraphics[width=0.33\textwidth]{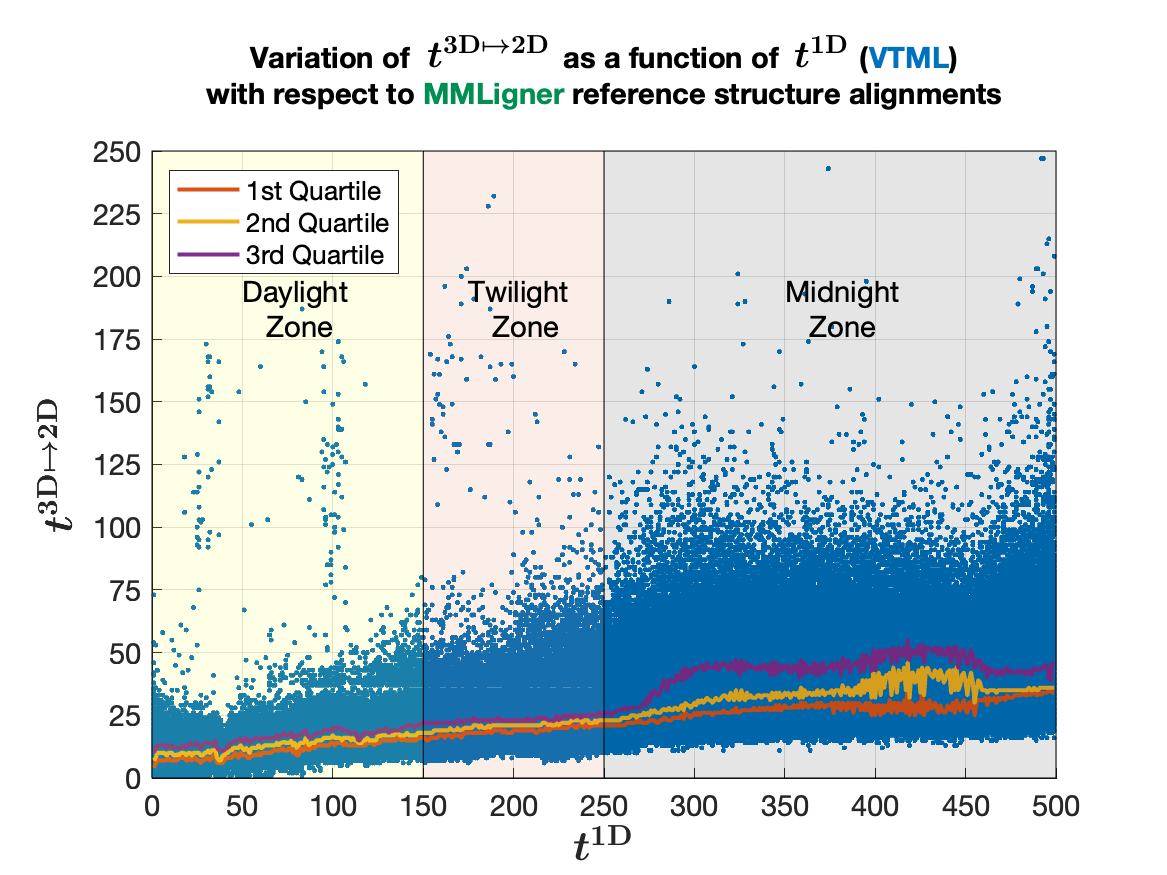} &
\includegraphics[width=0.33\textwidth]{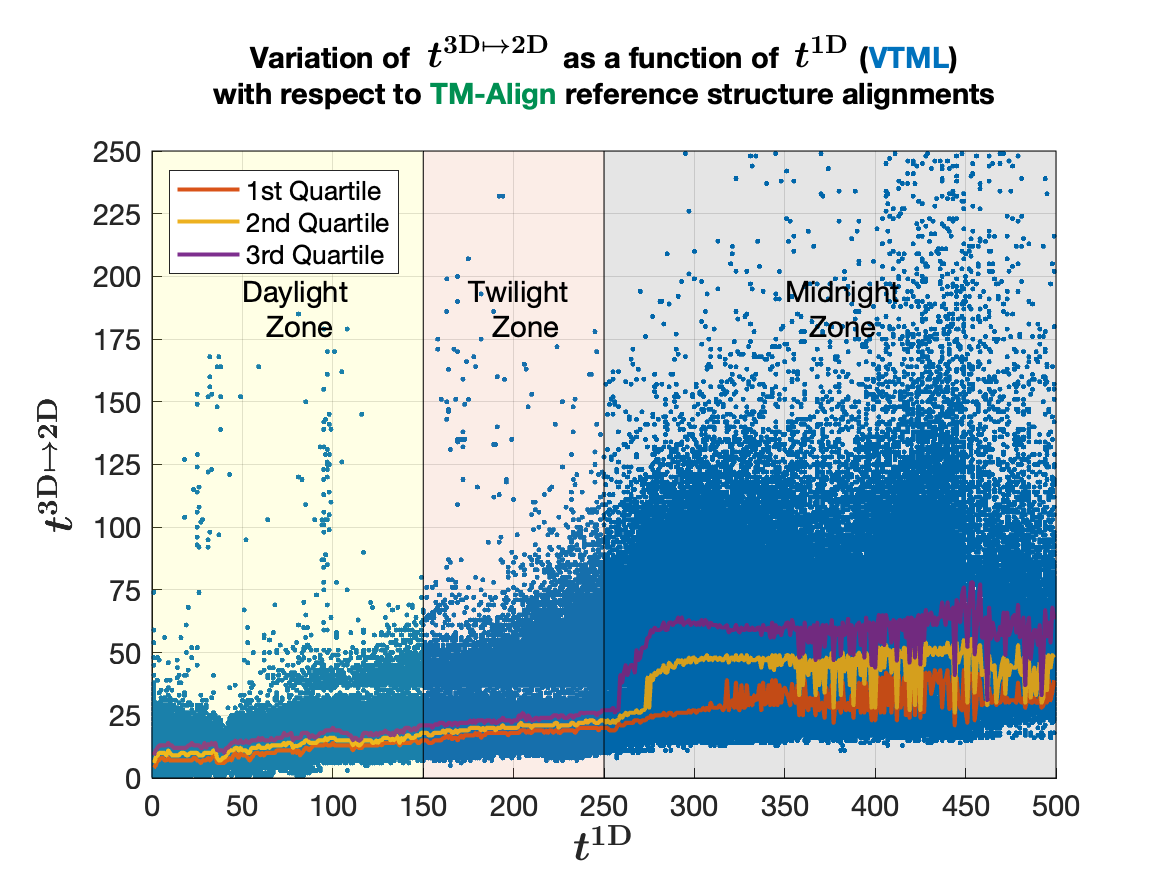} \\
\end{tabular}
\caption{Variation of the divergence times of structures ($\typeset{t}{3D$\mapsto$2D}$)  and sequences ($\typeset{t}{1D}$)  for million domain pairs. The three columns correspond to $\typeset{t}{3D$\mapsto$2D}$  (y-axes) inferred using 3 structure alignment programs: DALI, MMLigner, and TM-Align. The x-axes in each row from top to bottom correspond to $\typeset{t}{1D}$ inferred using 3 time-parameterized models: BLOSUM, MMLSUM, and VTML
and associated parameters respectively.} \label{fig:seqvsstrtime}
\end{figure*}

We used a dataset of 1 million domain pairs randomly sampled from family and superfamily levels of SCOP (v2.08) classification to generate three sets of structure alignment datasets ($\typeset{\mathbf{D}}{3D}$) from three alignment programs (1) MMLigner \citep{collier2017statistical} (2) TM-Align \citep{zhang2005tm}) and (3) DALI \citep{holm1995dali}. For each structure alignment in all three datasets, first, we inferred the divergence time of sequences $\typeset{t}{1D}$ using the SeqMMLigner  \citep{sumanaweera2019statistical} program.  We then selected the set of unique domains that constitute the million domain pairs datasets yielding a set of  147,018 unique domains. For each unique domain structure, we assigned the secondary structure states using the SST program \citep{konagurthu2012minimum} and we mapped each structure alignment in the three datasets mentioned previously, to its corresponding alignment of secondary structure states. This generated yet another three datasets of alignments ($\typeset{\mathbf{D}}{3D$\mapsto$2D}$) containing secondary structure states in
place of amino acid residues. Using these alignments, we inferred their respective divergence times $\typeset{t}{3D$\mapsto$2D}$ using SSTSUM (see Section \ref{sec:opttime}).

Fig. \ref{fig:seqvsstrtime} plots all possible variations involving the structure alignment datasets generated using DALI, MMLigner, and TM-Align (columns) and the stochastic models for amino acid change using BLOSUM~\citep{henikoff1992amino}, MMLSUM~\citep{sumanaweera2022bridging}, and VTML~\citep{muller2002estimating} (rows). The $x$ axis of each subplot shows the inferred $\typeset{t}{1D}$ for each alignment  in the specified structure alignment dataset (DALI, MMLigner, or TM-Align) using the specified amino acid stochastic model (BLOSUM, MMLSUM, or VTML) whereas the $y$ axis shows the inferred $\typeset{t}{3D$\mapsto$2D}$ of the same using SSTSUM.

Across all the subplots in Fig. \ref{fig:seqvsstrtime}, a similar trend is observed. All subplots have been demarcated into  `daylight' ($0 \leq \typeset{t}{1D} < 150$), `twilight'  ($150 \leq \typeset{t}{1D} < 250$), and `midnight' ($\typeset{t}{1D}\geq 250$) zones of sequence relationships~\citep{rajapaksa2022reliability}. As evident from the first (orange), second (yellow), and third (violet) quartile statistics on each subplot, it can be gathered that $\typeset{t}{3D$\mapsto$2D}$ increases
linearly in both daylight and twilight zones with respect to $\typeset{t}{1D}$. In the midnight zone, the relationship across various models of sequence time becomes noisy (at varying degrees across the plots). The high variation is a result of the amino acid sequence models in the midnight zone becoming unreliable in their inference of sequence relationships \citep{rajapaksa2022reliability}.

As a case study, consider the subplot corresponding to the MMLigner dataset using MMLSUM (subplot in the second row, the second column in Fig. \ref{fig:seqvsstrtime}), at $\typeset{t}{1D} = 150$ (indicating 72\% expected change of amino acids as per Fig. \ref{fig:expchange}), $\typeset{t}{3D$\mapsto$2D}$ ranges between $[16, 22]$ (indicating $[14, 19]\% $ expected change in secondary structure states) and at $\typeset{t}{1D} = 250$ (84\% expected change of amino acids), $\typeset{t}{3D$\mapsto$2D}$ ranges
between $[21, 26]$ ($[18, 21]\% $ expected change in secondary structure states).  An  interpretation of these statistics as if we were dealing with a  protein containing 100 residues diverging from its common ancestor is as follows.  If the protein is at the start of the twilight zone, the expectation is that $\sim72$ out of its 100 amino acids have changed, although its secondary structure states compared to that ancestral protein is expected to have undergone $\sim14$-$19$ changes. This increases to $\sim 18$-$21$ if the protein relationship were at the start of the midnight zone where $\sim84$ amino acids are expected to change. This is consistent with the general observation that tertiary structures (and hence secondary)  change more conservatively than their amino acid sequences but is made here more quantitatively precise in terms of the relationship we can gather from  time $t^{(\text{1D})}$. From looking at the quartile statistics, especially in the daylight and twilight zones across
all plots, the sequence time is $\sim 9.6\times$ greater than the structure time, and the accumulated change in a sequence is $\sim 4.5\times$ more than in a structure. This quantifies the relationship between the divergence of sequence with the divergence of structure.

\begin{figure*}[htb]
\begin{tabular}{@{}c@{}c@{}c@{}}
    \includegraphics[width=0.33\textwidth]{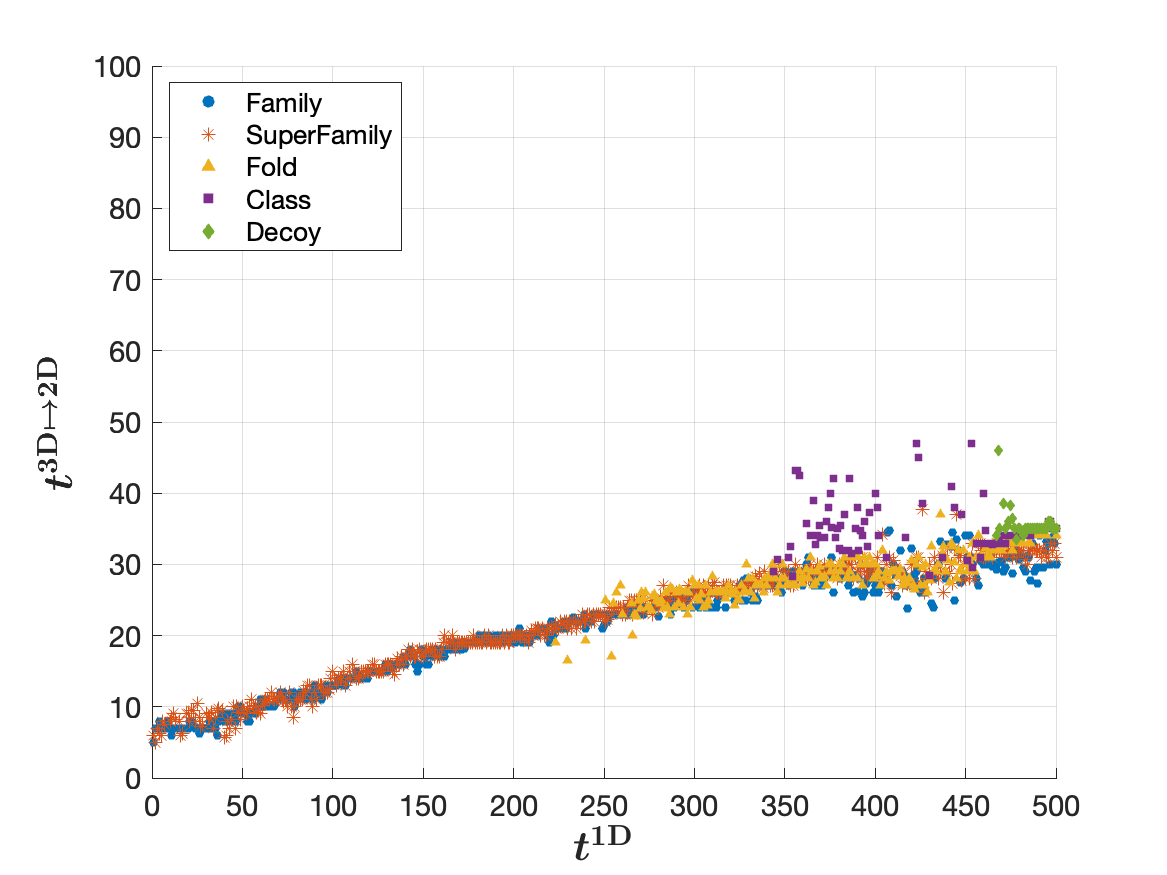} & \includegraphics[width=0.33\textwidth]{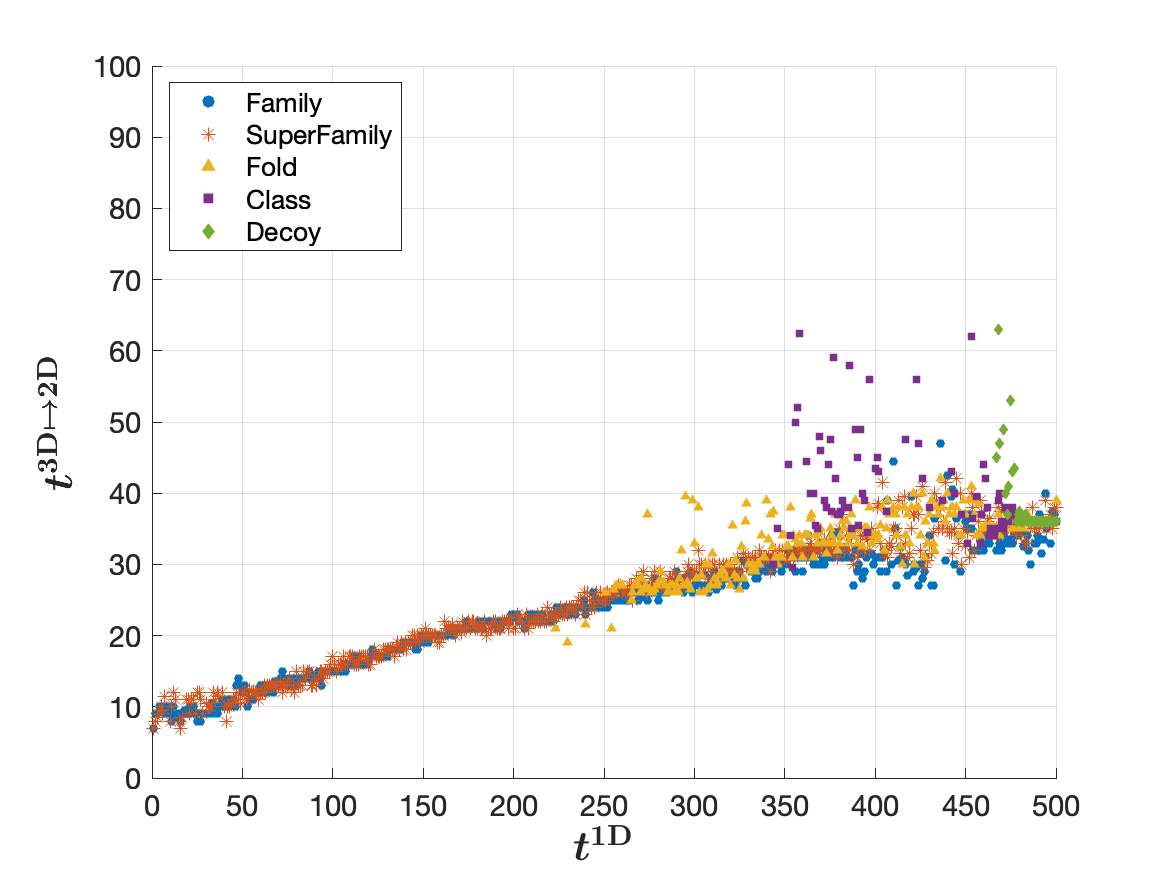} & \includegraphics[width=0.33\textwidth]{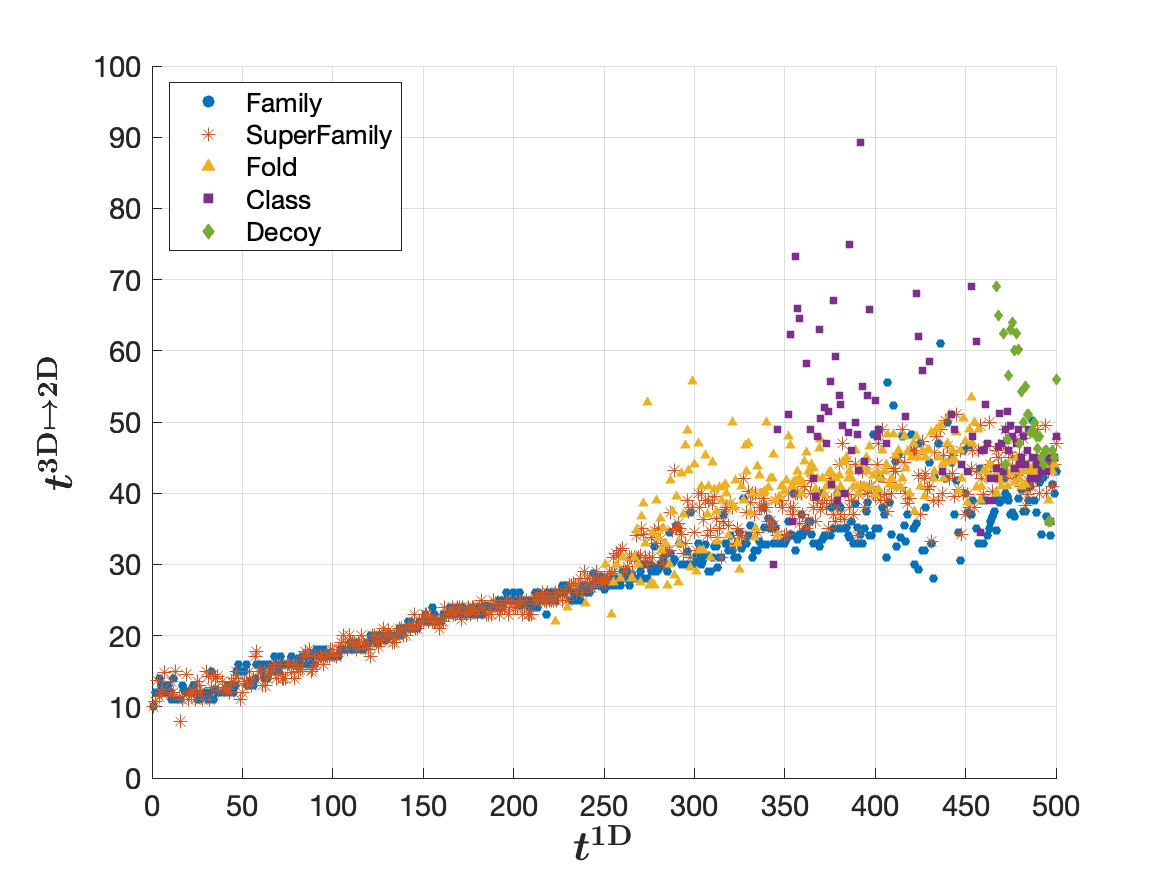} \\
    (a) & (b) & (c)
\end{tabular}
\caption{Variation of the (a) 1st (b) 2nd (c) 3rd quartile statistics of divergence times of structures ($\typeset{t}{3D$\mapsto$2D}$) and sequences ($\typeset{t}{1D}$) for domain-pairs drawn from 5 different levels of SCOP classification. In all three plots, blue circles, red stars, yellow triangles, purple squares, and green diamonds denote domain pairs sampled from family, superfamily, fold, class, and two different class levels of SCOP, respectively.} \label{fig:seqvsstrtimescoplevels}
\end{figure*}

We conduct another analysis to compare the divergence time of structures based on the structural distances implicit in the hierarchical levels of SCOP. Through a random sampling process (followed by filtering) we derived 55,201 domain pairs from the same family level, 31,600 from the same superfamily level, 40,582 from the same fold level, 40,551 from the same class level, and 40,466 from different classes (referred to as decoy). We then inferred $\typeset{t}{3D$\mapsto$2D}$ of those pairs after aligning their structures using MMLigner. We also inferred $\typeset{t}{1D}$ using SeqMMLigner for each structure alignment in these 5 domain-pairs datasets. Fig. \ref{fig:seqvsstrtimescoplevels} plots the variation of the quartile statistics of $\typeset{t}{3D$\mapsto$2D}$ as a function of $\typeset{t}{1D}$.

The divergence time $\typeset{t}{3D$\mapsto$2D}$ of domain pairs that belong to family and superfamily levels increases linearly with $\typeset{t}{1D}$ in the range $[0, \sim 170]$. The slope reduces slightly thereafter for $\typeset{t}{1D} \geq 170$, which can be interpreted as the structures continuing to be conserved even when the sequences diverge into the twilight zone of sequence relationships. The divergence time $\typeset{t}{3D$\mapsto$2D}$ of domain pairs at the superfamily level is slightly higher than that of the family level after $\typeset{t}{1D} \geq 250$ (midnight zone), which is expected as the superfamily relationships are distant in sequence with structures maintaining the core with variations in the peripheral regions of the fold ~\citep{chothia1986relation}. On average, the structure time of domain pairs that belong to the fold level follows a similar trend except for some outliers. The outliers suggest that while the folding pattern is preserved, parts of the structures have deviated into different conformations. Further, an explosion in  $\typeset{t}{3D$\mapsto$2D}$ of domain pairs in the remaining two sets (class and decoy) is  observed. This is
consistent with the observation that proteins that share relationships no better than the level of class or worse (fall in different classes) as per SCOP have completely different structures, and hence are unrelated. This gives a consistency test of our time-parameterized models with the notion of sequence and structure distance internalized in the SCOP hierarchy.

\begin{table*}[htb]
\footnotesize
\caption{Divergence times and the expected changes inferred based on SSTSUM (secondary structure) and MMLSUM(amino acid), comparing the structures and  sequences of a set of five  homologous domains from the Globin-like folds with the $\alpha$ chain of human hemoglobin ($\mathtt{1HHO}$ chain $\mathtt{A}$).  }
\label{tbl:globins}
\begin{tabular}{ l C{1.2cm} C{1.4cm} C{1.2cm} C{1.3cm} C{1.2cm} C{1.4cm}  }
    \hline
    $S$ vs  $T$   & $\typeset{t}{3D$\mapsto$2D}$  & Exp. s.s. change & RMSD  & $N_{\text{equivalence}}$ & $\typeset{t}{1D}$ & Exp. a.a. change  \\ \hline\hline
    human (1HHO(A)) vs chicken (1HBR(A)) & 16 & 14.4\% & 0.9\AA & 140 &  52  & 39.1\% \\
    human (1HHO(A)) vs sperm whale (1MBD) & 17 & 15.1\%  & 1.5\AA & 140 & 156 & 73.2\% \\
    human (1HHO(A)) vs \textit{Chironomus} (1ECA(A)) & 18 & 15.9\% & 2.5\AA & 131 & 200 & 79.7\%\\
    human (1HHO(A)) vs bacterium (4VHB(A)) & 18 & 15.9\% & 2.1\AA & 128 & 221 & 81.9\%\\
    human (1HHO(A)) vs red-alga  (2BV8(A)) & 23 & 19.4\% & 4.0\AA & 126 & 274 & 86.0\% \\
    \hline
\end{tabular}
\end{table*}
\begin{figure*}[h!]
\begin{tabular}{@{}c@{}c@{}c@{}c@{}c@{}}
    \includegraphics[width=0.175\textwidth]{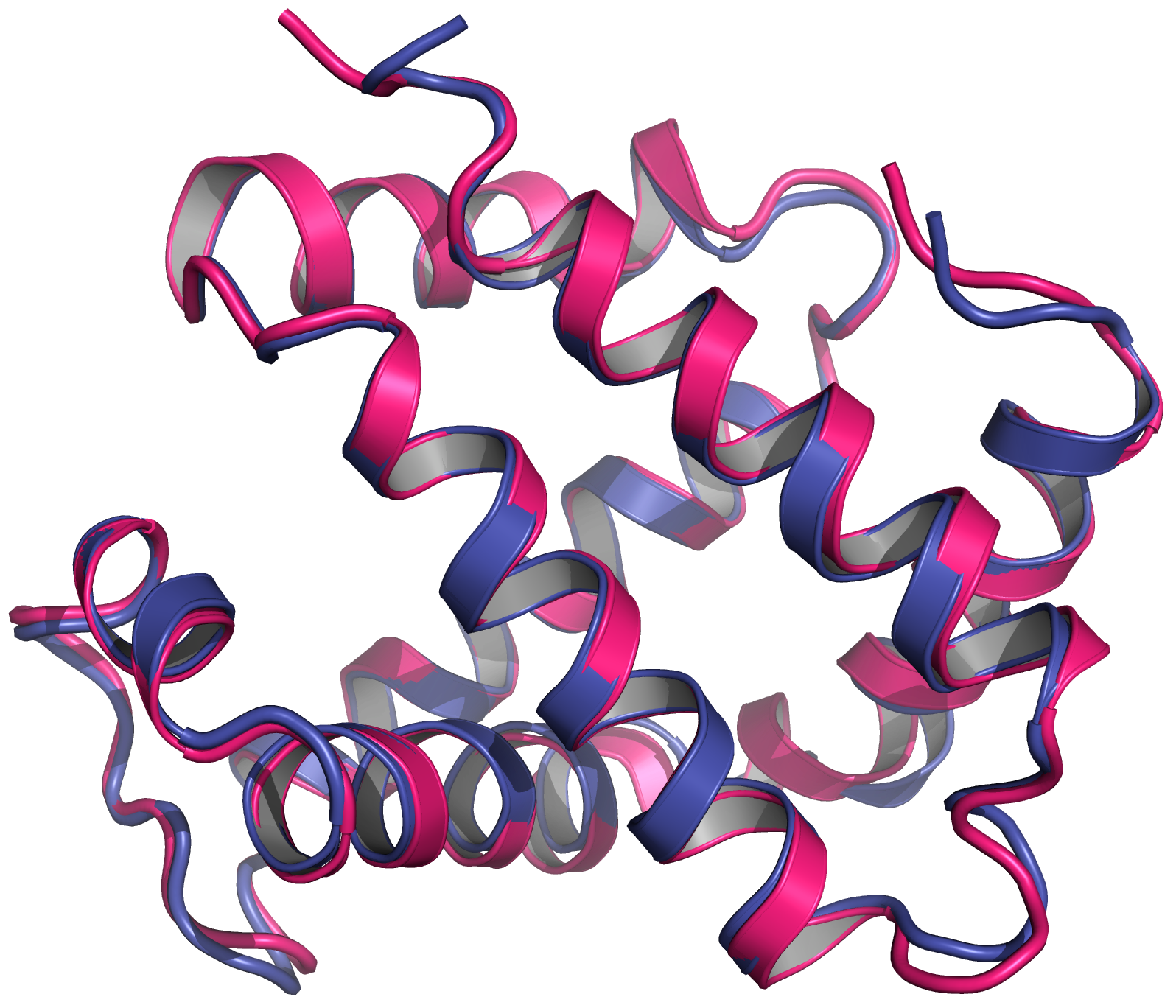} &
    \includegraphics[width=0.175\textwidth]{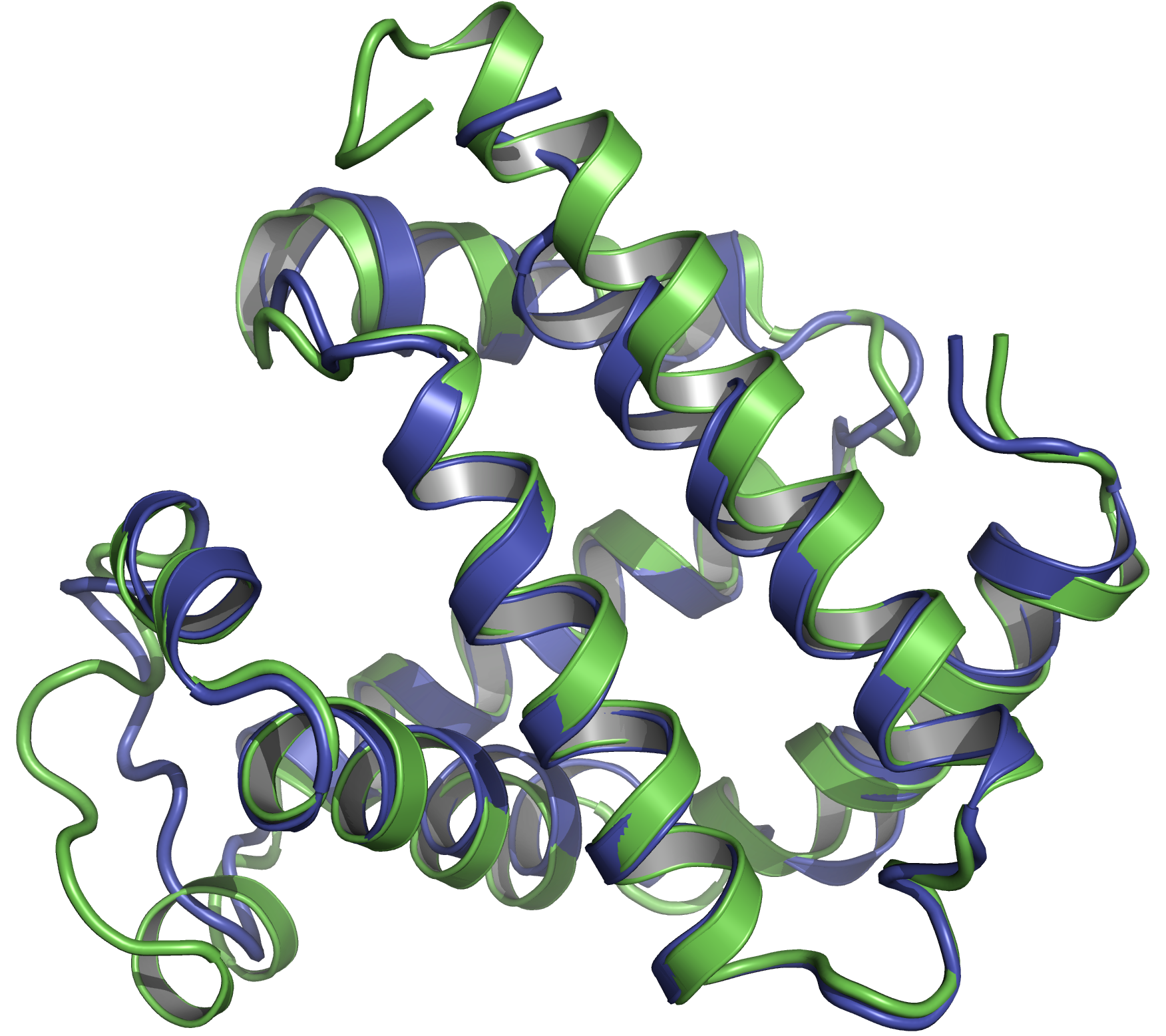} &
    \includegraphics[width=0.175\textwidth]{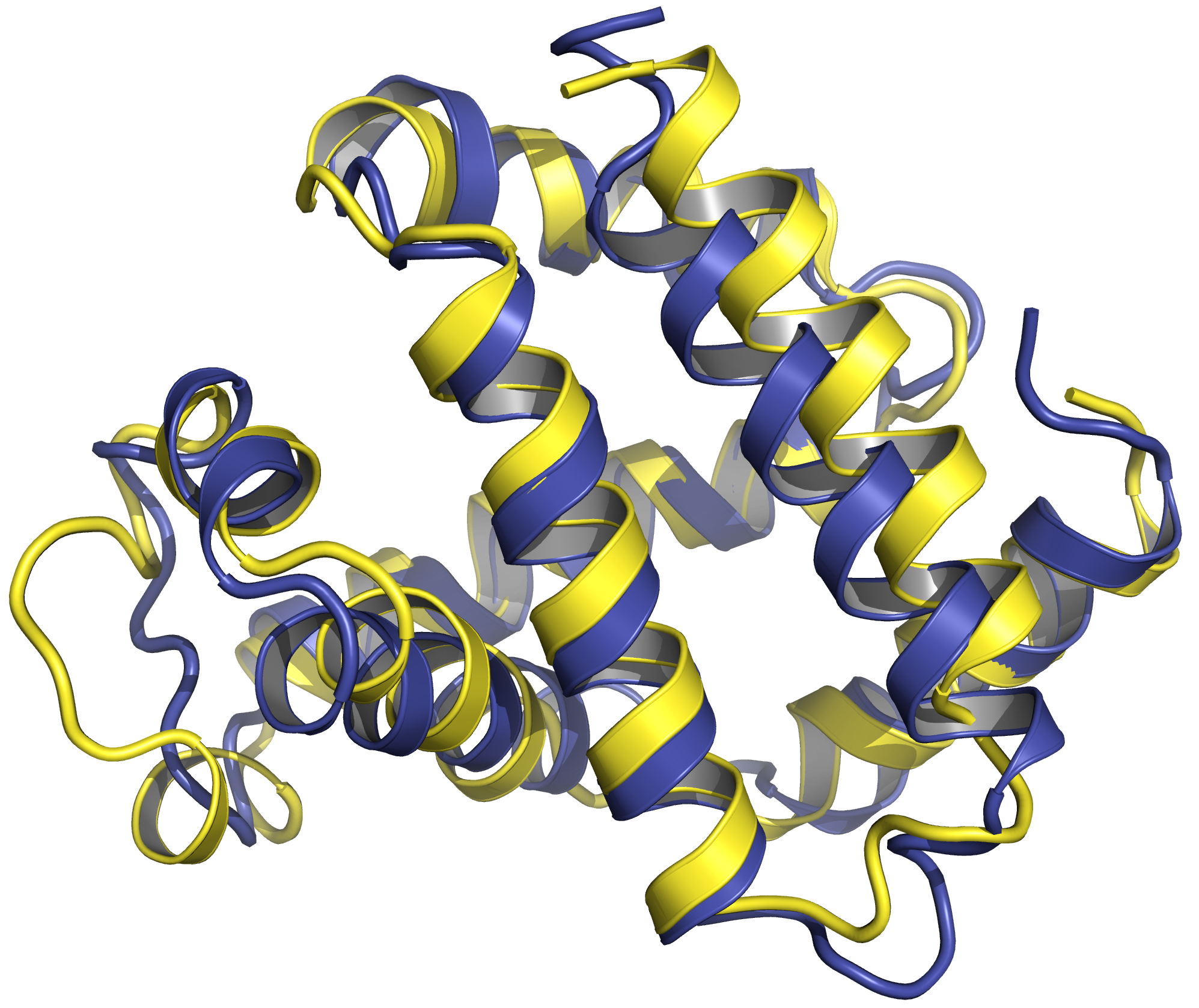}  &
    \includegraphics[width=0.16\textwidth]{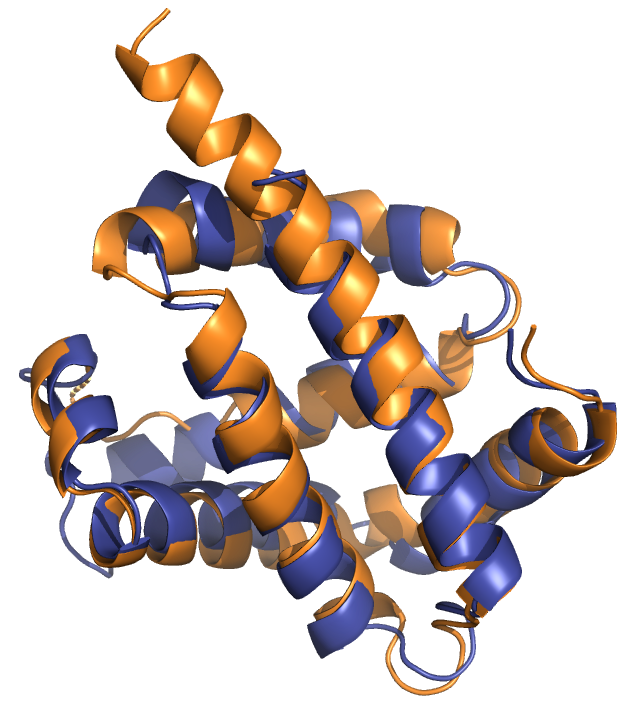} &
    \includegraphics[width=0.215\textwidth]{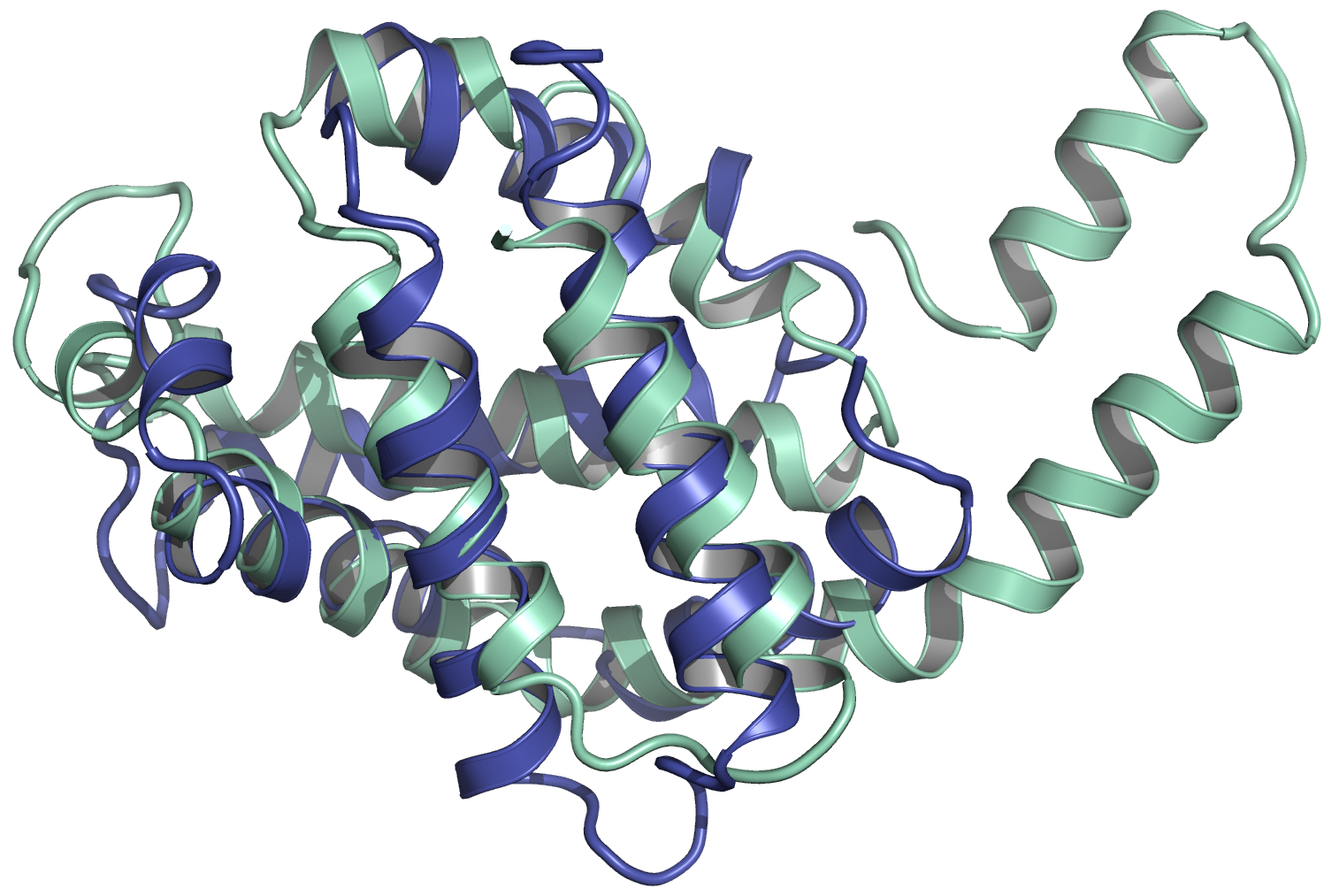} \\
    $\mathtt{1HHO(A)}$ vs $\mathtt{1HBR(A)}$ & $\mathtt{1HHO(A)}$ vs $\mathtt{1MBD}$  & $\mathtt{1HHO(A)}$ vs $\mathtt{1ECA(A)}$  & $\mathtt{1HHO(A)}$ vs $\mathtt{4VHB(A)}$  & $\mathtt{1HHO(A)}$ vs $\mathtt{2BV8(A)}$  \\
\end{tabular}
\caption{Superpositions of the structures of the five  homologous domains with the $\alpha$ chain of human hemoglobin.} \label{fig:casestudy}
\end{figure*}

\subsection{A case study of globin-like folds}

We present a qualitative case study comparing both the structure and sequence of human hemoglobin ($\mathtt{1HHO}$ chain $\mathtt{A}$) with five other related but an evolutionarily diverging set of proteins: chicken hemoglobin ($\mathtt{1HBR}$ chain $\mathtt{A}$), sperm whale myoglobin ($\mathtt{1MBD}$), \textit{Chironomus} erythrocruorin ($\mathtt{1ECA}$ chain $\mathtt{A}$), bacterial hemoglobin ($\mathtt{4VHB}$ chain $\mathtt{A}$), and red-alga phycocyanin ($\mathtt{2BV8}$ chain $\mathtt{A}$), all classified within the same fold in SCOP.

Table \ref{tbl:globins} presents the divergence times and the corresponding expected changes (see Equation \ref{eq:expchange}) inferred based on SSTSUM (secondary structure) and MMLSUM (amino acid) for the 5 pairs of proteins. To visually support these statistics, the superpositions of the structures of each pair (using MMLigner) are shown in Fig. \ref{fig:casestudy}. The rows of table \ref{tbl:globins} appear in the increasing order of the divergence time of sequences ($\typeset{t}{1D}$). The
change in amino acids between the sequences of chicken and human hemoglobin  is expected to be at 39.1\%. This change increases to 73.2\% for human hemoglobin vs sperm whale myoglobin and to 79.7\% for human hemoglobin vs \textit{Chironomus} erythrocruorin. The sequences of human and bacterial hemoglobin are expected to undergo 81.9\% change as shown in Table \ref{tbl:globins}. But, the structures of these 4 pairs are expected to undergo only a slight change from 14.4\% to $15.9\%$ of secondary structures. Also notice that  RMSD is less reliable to track such a change effectively, as it is always a function of the number of equivalences/correspondences in an alignment. In other words, this measure could yield imprecise conclusions, with lower RMSD values for more distantly related pairs of proteins than their closer counterparts, merely by trading off the number of equivalences in the alignment (refer rows 3-4 of Table \ref{tbl:globins}). All these proteins including human
hemoglobin (except for phycocyanin) show a `positive cooperativity' with oxygen binding, a very critical function to preserve under evolutionary pressure \citep{knapp1999structural,steigemann1979structure,bolognesi1999anticooperative}.
As highlighted by \citet{pastore1990comparison}, globins and phycocyanins have a sequence similarity in the midnight zone (86\% change in amino acid sequence w.r.t. human hemoglobin), fit poorly when superposed at the level of their central $\alpha$-Carbon atoms (RMSD 4.0\AA  over 126 equivalences with $\mathtt{1HHO}$ ($\mathtt{A}$)), yet have a folding pattern  similar to globins (with estimated 19.4\% change in secondary structure states).
Thus, the statistics in Table \ref{tbl:globins} support the
observations of \citet{pastore1990comparison}.

\subsection{Application in secondary structure prediction}

Beyond the implication of the relationship between the divergence of structures and divergence of sequences to study the evolution of protein domains, we demonstrate here an application of the time-parameterized SSTSUM models and its relationship with amino acid time that we inferred above.

In particular, we used SSTSUM and associated models and tested its utility for secondary structure prediction, a task that takes a query sequence to predict  3 states of secondary structures. Given a query sequence, we first search for a set of local alignments (hits) in a non-redundant dataset containing 45,887 protein sequences that were deposited before $1^{st}$ of January 2017 (see supplementary materials for the PDB IDs).  We emphasize that we only use these sequences and their
known secondary structures, even though other programs use much larger data sets, often containing information of more than 10 million protein sequences.

For each hit from this set, we infer $\typeset{t}{1D}$ using its sequence alignments with local regions of the query. Finally, we use SSTSUM to  estimate the conditional probabilities of each query being in one of the three secondary structure states using the method explained in Section \ref{sect:predframework}, and choose the state with the highest of the three probabilities to predict the secondary structure at each position of the query.

We evaluate this method of prediction with 3 other widely used secondary structure predictors that all use neural architecture: Deep-CNF \citep{wang2016protein}, PSIPRED \citep{mcguffin2000psipred}, and JPred \citep{drozdetskiy2015jpred4} using the targets in the recent series of community-wide experiments on Critical Assessment of Protein Structure Prediction (CASP) 14, and 15.

Note that the dataset on which we build the notion of \emph{hits} has structures submitted before CASP14 (in 2020) and CASP15 (in 2022) was held. By doing this, we can ensure we do not `benefit' from \emph{leakage}~\citep{gibney2022ai,kapoor2022leakage} of information during our prediction experiment. However, the same cannot be ascertained for the predictors we are comparing against. Yet we proceed as if these predictors we are comparing do not have any leakage from their training set
into their testing/prediction.

\begin{figure*}[h]
\begin{tabular}{@{}c@{}c@{}}
    \includegraphics[width=0.5\textwidth]{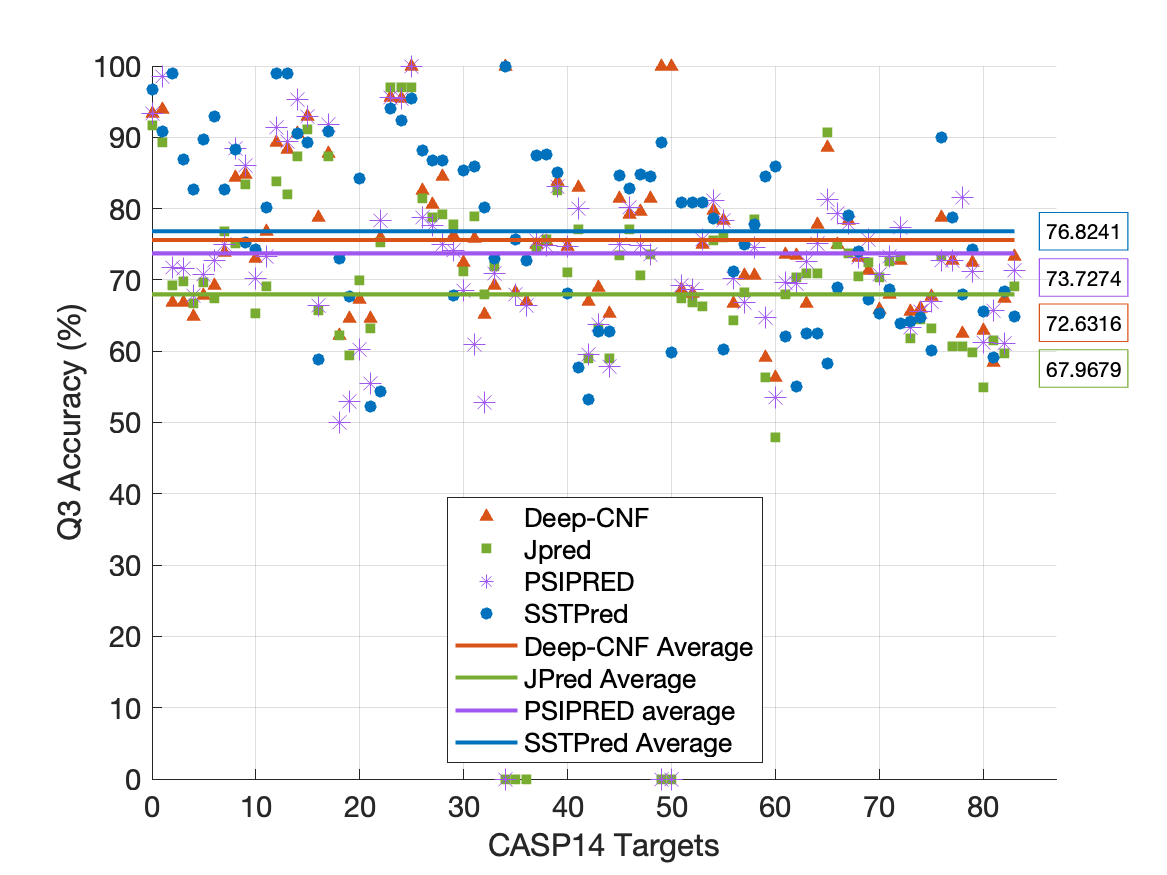} & \includegraphics[width=0.5\textwidth]{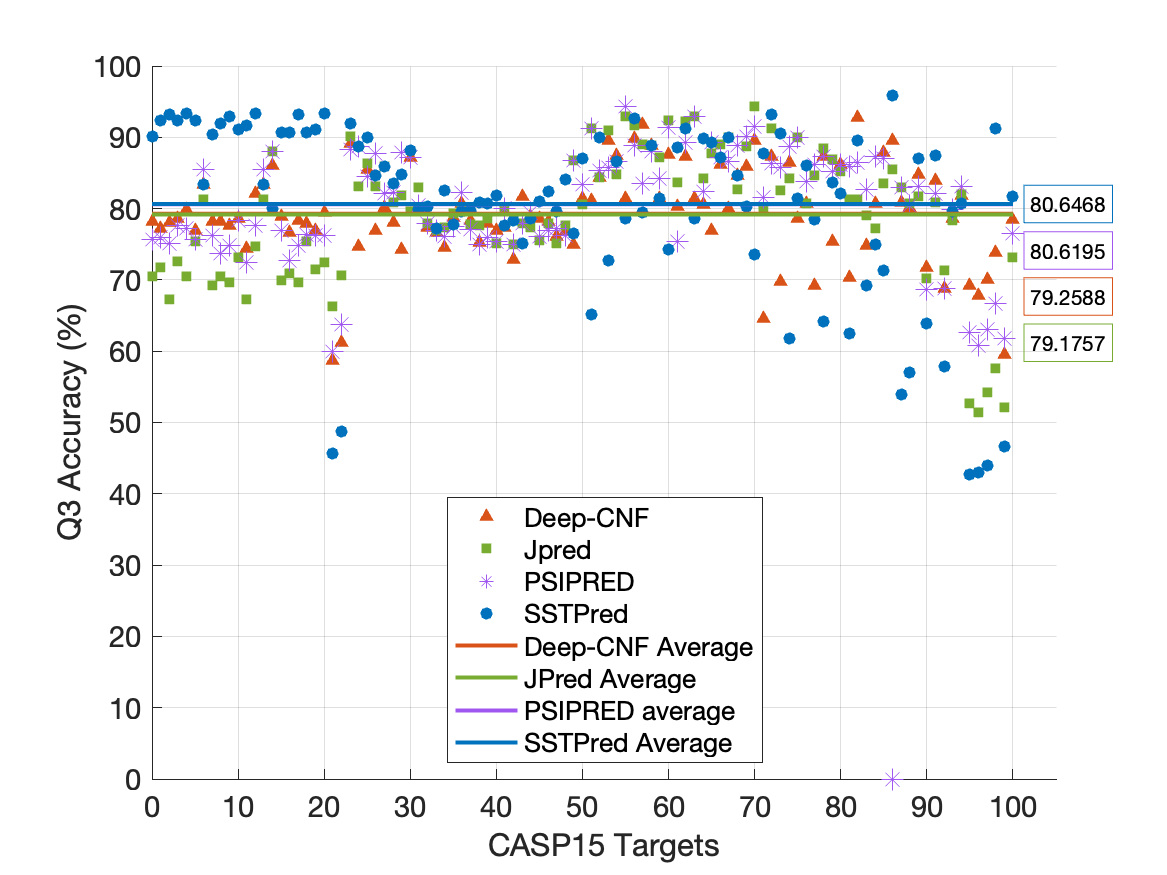} \\
    (a) & (b)
\end{tabular}
\caption{Q3 accuracies of CASP14 (left plot) and CASP15 (right plot) targets considering secondary structure assignments produced by SST \citep{konagurthu2012minimum} as reference. In both plots,  red triangles correspond to accuracies of the predictions produced by Deep-CNF, green squares to those of JPred, purple stars to those of PSIPRED, and blue points to those of our method SSTPred. The average of each method is shown as a colored line.}
\label{fig:q3acc}
\end{figure*}

To evaluate our method's prediction performance, we employ Q3 accuracy, which measures the percentage of correctly predicted residues for the 3 states of secondary structures with respect to a reference secondary structure  assignment. Fig.  \ref{fig:q3acc} shows the Q3 accuracies of all four predictors on the CASP 14 and 15 datasets with respect to the secondary structure  assignments produced by the SST program \citep{konagurthu2012minimum}. The PDB IDs of the targets in both CASP14 and CASP15 are available in the supplementary material. The results show that our simple method of prediction (SSTPRED) benefiting from SSTSUM models is competitive with the methods that apply neural architectures with extensive training and parameters. This supports the quality of the SSTSUM models we have inferred.

\section{Conclusion}
We  infer a time-parameterized Markov matrix SSTSUM which models the changes to structure through patterns of conservation of its secondary structure states. Furthermore, we infer associated time-parameterized Dirichlet parameters that work jointly with SSTSUM to model insertion and deletion events. This combined model allows the inference of the divergence time of protein structures. We have used SSTSUM to analyze the  Markov time of divergence of one million domain pairs.
This has uniquely allowed us to relate the divergence time of sequences with structures. This analysis shows that the sequence time runs  $\sim 9.6\times$ faster than structure time, and the accumulated changes to the sequence is $\sim 4.5\times$ more than to the structure, thus  quantifying the otherwise general observation that sequences change more drastically than structures in evolution. Complementing this observation is another analysis we performed to compare the divergence time of structures and sequences in the hierarchical levels of SCOP. The inferred relationship correlates well with the notion of sequence and structure distance implicit in the SCOP hierarchy.

Finally, we demonstrate a potential application of the results of our work in the form of secondary structure prediction. Although the method we used for prediction was simple and based on standard statistics, the use of SSTSUM and the relationship between the divergence times of sequence and (secondary) structure, yields a performance that stands competitive with other prediction methods. As a future work, it would be interesting to examine the implication of these models in the
construction of evolutionary trees, among other applications.

All models inferred in this study (along with the programs and raw data) are available on \url{https://lcb.infotech.monash.edu/sstsum}.

\bibliographystyle{natbib}
\bibliography{paper}

\begin{thebibliography}{}

\bibitem[Allison(2018)Allison]{allison2018coding}
Allison, L. (2018).
\newblock {\em {Coding Ockham's Razor}\/}.
\newblock Springer.

\bibitem[Bolognesi {\em et~al.}(1999)Bolognesi, Boffi, Coletta, Mozzarelli,
  Pesce, Tarricone, and Ascenzi]{bolognesi1999anticooperative}
Bolognesi, M.  {\em et~al.} (1999).
\newblock Anticooperative ligand binding properties of recombinant ferric
  {Vitreoscilla} homodimeric hemoglobin: a thermodynamic, kinetic and {X}-ray
  crystallographic study.
\newblock {\em Journal of Molecular Biology\/}, {\bf 291}(3), 637--650.

\bibitem[Bromham and Penny(2003)Bromham and Penny]{bromham2003modern}
Bromham, L. and Penny, D. (2003).
\newblock The modern molecular clock.
\newblock {\em Nature Reviews Genetics\/}, {\bf 4}(3), 216--224.

\bibitem[Chothia and Lesk(1986)Chothia and Lesk]{chothia1986relation}
Chothia, C. and Lesk, A.~M. (1986).
\newblock The relation between the divergence of sequence and structure in
  proteins.
\newblock {\em The EMBO Journal\/}, {\bf 5}(4), 823--826.

\bibitem[Collier {\em et~al.}(2017)Collier, Allison, Lesk, Stuckey, Garcia
  de~la Banda, and Konagurthu]{collier2017statistical}
Collier, J.~H.  {\em et~al.} (2017).
\newblock Statistical inference of protein structural alignments using
  information and compression.
\newblock {\em Bioinformatics\/}, {\bf 33}(7), 1005--1013.

\bibitem[Dayhoff {\em et~al.}(1978)Dayhoff, Schwartz, and Orcutt]{dayhoff1978}
Dayhoff, M.  {\em et~al.} (1978).
\newblock A model of evolutionary change in proteins.
\newblock {\em Atlas of Protein Sequence and Structure\/}, {\bf 5}, 345--352.

\bibitem[Dempster {\em et~al.}(1977)Dempster, Laird, and
  Rubin]{dempster1977maximum}
Dempster, A.~P.  {\em et~al.} (1977).
\newblock Maximum likelihood from incomplete data via the {EM} algorithm.
\newblock {\em Journal of the Royal Statistical Society: Series B
  (Methodological)\/}, {\bf 39}(1), 1--22.

\bibitem[Drozdetskiy {\em et~al.}(2015)Drozdetskiy, Cole, Procter, and
  Barton]{drozdetskiy2015jpred4}
Drozdetskiy, A.  {\em et~al.} (2015).
\newblock {JPred4}: a protein secondary structure prediction server.
\newblock {\em Nucleic Acids Research\/}, {\bf 43}(W1), W389--W394.

\bibitem[Echave {\em et~al.}(2016)Echave, Spielman, and
  Wilke]{echave2016causes}
Echave, J.  {\em et~al.} (2016).
\newblock Causes of evolutionary rate variation among protein sites.
\newblock {\em Nature Reviews Genetics\/}, {\bf 17}(2), 109--121.

\bibitem[Gibney(2022)Gibney]{gibney2022ai}
Gibney, E. (2022).
\newblock Is {AI} fuelling a reproducibility crisis in science?
\newblock {\em Nature\/}, {\bf 608}, 250--251.

\bibitem[Gonnet {\em et~al.}(1992)Gonnet, Cohen, and
  Benner]{gonnet1992exhaustive}
Gonnet, G.~H.  {\em et~al.} (1992).
\newblock Exhaustive matching of the entire protein sequence database.
\newblock {\em Science\/}, {\bf 256}(5062), 1443--1445.

\bibitem[Henikoff and Henikoff(1992)Henikoff and Henikoff]{henikoff1992amino}
Henikoff, S. and Henikoff, J.~G. (1992).
\newblock Amino acid substitution matrices from protein blocks.
\newblock {\em Proceedings of the National Academy of Sciences\/}, {\bf
  89}(22), 10915--10919.

\bibitem[Holm and Sander(1995)Holm and Sander]{holm1995dali}
Holm, L. and Sander, C. (1995).
\newblock {Dali}: a network tool for protein structure comparison.
\newblock {\em Trends in Biochemical Sciences\/}, {\bf 20}(11), 478--480.

\bibitem[Holmes(1998)Holmes]{holmes1998studies}
Holmes, I. (1998).
\newblock Studies in probabilistic sequence alignment and evolution.
\newblock {\em Queens' College\/}.

\bibitem[Kapoor and Narayanan(2022)Kapoor and Narayanan]{kapoor2022leakage}
Kapoor, S. and Narayanan, A. (2022).
\newblock {Leakage and the Reproducibility Crisis in ML-based Science}.
\newblock {\em arXiv preprint arXiv:2207.07048\/}.

\bibitem[Kinch and Grishin(2002)Kinch and Grishin]{kinch2002evolution}
Kinch, L.~N. and Grishin, N.~V. (2002).
\newblock Evolution of protein structures and functions.
\newblock {\em Current Opinion in Structural Biology\/}, {\bf 12}(3), 400--408.

\bibitem[Knapp {\em et~al.}(1999)Knapp, Oliveira, Xie, Ernst, Riggs, and
  Hackert]{knapp1999structural}
Knapp, J.~E.  {\em et~al.} (1999).
\newblock The structural and functional analysis of the hemoglobin {D}
  component from chicken.
\newblock {\em Journal of Biological Chemistry\/}, {\bf 274}(10), 6411--6420.

\bibitem[Konagurthu {\em et~al.}(2006)Konagurthu, Whisstock, Stuckey, and
  Lesk]{konagurthu2006mustang}
Konagurthu, A.~S.  {\em et~al.} (2006).
\newblock {MUSTANG}: a multiple structural alignment algorithm.
\newblock {\em Proteins: Structure, Function, and Bioinformatics\/}, {\bf
  64}(3), 559--574.

\bibitem[Konagurthu {\em et~al.}(2012)Konagurthu, Lesk, and
  Allison]{konagurthu2012minimum}
Konagurthu, A.~S.  {\em et~al.} (2012).
\newblock Minimum message length inference of secondary structure from protein
  coordinate data.
\newblock {\em Bioinformatics\/}, {\bf 28}(12), i97--i105.

\bibitem[McGuffin {\em et~al.}(2000)McGuffin, Bryson, and
  Jones]{mcguffin2000psipred}
McGuffin, L.~J.  {\em et~al.} (2000).
\newblock The {PSIPRED} protein structure prediction server.
\newblock {\em Bioinformatics\/}, {\bf 16}(4), 404--405.

\bibitem[M{\"u}ller {\em et~al.}(2002)M{\"u}ller, Spang, and
  Vingron]{muller2002estimating}
M{\"u}ller, T.  {\em et~al.} (2002).
\newblock Estimating amino acid substitution models: a comparison of
  {Dayhoff's} estimator, the resolvent approach and a maximum likelihood
  method.
\newblock {\em Molecular Biology and Evolution\/}, {\bf 19}(1), 8--13.

\bibitem[Murzin {\em et~al.}(1995)Murzin, Brenner, Hubbard, and
  Chothia]{murzin1995scop}
Murzin, A.~G.  {\em et~al.} (1995).
\newblock {SCOP:} a structural classification of proteins database for the
  investigation of sequences and structures.
\newblock {\em Journal of Molecular Biology\/}, {\bf 247}(4), 536--540.

\bibitem[P{\'a}l {\em et~al.}(2006)P{\'a}l, Papp, and
  Lercher]{pal2006integrated}
P{\'a}l, C.  {\em et~al.} (2006).
\newblock An integrated view of protein evolution.
\newblock {\em Nature Reviews Genetics\/}, {\bf 7}(5), 337--348.

\bibitem[Pastore and Lesk(1990)Pastore and Lesk]{pastore1990comparison}
Pastore, A. and Lesk, A.~M. (1990).
\newblock Comparison of the structures of globins and phycocyanins: evidence
  for evolutionary relationship.
\newblock {\em Proteins: Structure, Function, and Bioinformatics\/}, {\bf
  8}(2), 133--155.

\bibitem[Rajapaksa {\em et~al.}(2022)Rajapaksa, Sumanaweera, Lesk, Allison,
  Stuckey, Garcia de~la Banda, Abramson, and
  Konagurthu]{rajapaksa2022reliability}
Rajapaksa, S.  {\em et~al.} (2022).
\newblock On the reliability and the limits of inference of amino acid sequence
  alignments.
\newblock {\em Bioinformatics\/}, {\bf 38}(Supplement\_1), i255--i263.

\bibitem[Sarich and Wilson(1967)Sarich and Wilson]{sarich1967immunological}
Sarich, V.~M. and Wilson, A.~C. (1967).
\newblock Immunological time scale for hominid evolution.
\newblock {\em Science\/}, {\bf 158}(3805), 1200--1203.

\bibitem[Soskine and Tawfik(2010)Soskine and Tawfik]{soskine2010mutational}
Soskine, M. and Tawfik, D.~S. (2010).
\newblock Mutational effects and the evolution of new protein functions.
\newblock {\em Nature Reviews Genetics\/}, {\bf 11}(8), 572--582.

\bibitem[Steigemann and Weber(1979)Steigemann and
  Weber]{steigemann1979structure}
Steigemann, W. and Weber, E. (1979).
\newblock Structure of erythrocruorin in different ligand states refined at
  1{\textperiodcentered} 4 {\aa} resolution.
\newblock {\em Journal of Molecular Biology\/}, {\bf 127}(3), 309--338.

\bibitem[Sumanaweera {\em et~al.}(2019)Sumanaweera, Allison, and
  Konagurthu]{sumanaweera2019statistical}
Sumanaweera, D.  {\em et~al.} (2019).
\newblock Statistical compression of protein sequences and inference of
  marginal probability landscapes over competing alignments using finite state
  models and {Dirichlet} priors.
\newblock {\em Bioinformatics\/}, {\bf 35}(14), i360--i369.

\bibitem[Sumanaweera {\em et~al.}(2022)Sumanaweera, Allison, and
  Konagurthu]{sumanaweera2022bridging}
Sumanaweera, D.  {\em et~al.} (2022).
\newblock Bridging the gaps in statistical models of protein alignment.
\newblock {\em Bioinformatics\/}, {\bf 38}(Supplement\_1), i229--i237.

\bibitem[Wallace(2005)Wallace]{wallace2005statistical}
Wallace, C.~S. (2005).
\newblock {\em Statistical and inductive inference by {Minimum Message
  Length}\/}.
\newblock Springer Science \& Business Media.

\bibitem[Wang {\em et~al.}(2016)Wang, Peng, Ma, and Xu]{wang2016protein}
Wang, S.  {\em et~al.} (2016).
\newblock Protein secondary structure prediction using deep convolutional
  neural fields.
\newblock {\em Scientific Reports\/}, {\bf 6}(1), 1--11.

\bibitem[Worth {\em et~al.}(2009)Worth, Gong, and
  Blundell]{worth2009structural}
Worth, C.~L.  {\em et~al.} (2009).
\newblock Structural and functional constraints in the evolution of protein
  families.
\newblock {\em Nature Reviews Molecular Cell Biology\/}, {\bf 10}(10),
  709--720.

\bibitem[Zhang and Skolnick(2005)Zhang and Skolnick]{zhang2005tm}
Zhang, Y. and Skolnick, J. (2005).
\newblock {TM-align}: a protein structure alignment algorithm based on the
  {TM-score}.
\newblock {\em Nucleic Acids Research\/}, {\bf 33}(7), 2302--2309.

\bibitem[Zuckerkandl and Pauling(1965)Zuckerkandl and
  Pauling]{zuckerkandl1965evolutionary}
Zuckerkandl, E. and Pauling, L. (1965).
\newblock Evolutionary divergence and convergence in proteins.
\newblock In {\em Evolving Genes and Proteins\/}, pages 97--166. Elsevier.

\end{thebibliography}
\end{document}